\documentclass[utf8]{FrontiersinHarvard} 

\usepackage{url,hyperref,lineno,microtype,subcaption}
\usepackage[onehalfspacing]{setspace}
\usepackage{placeins}
\usepackage{ulem}

\def\keyFont{\fontsize{8}{11}\helveticabold }
\def\firstAuthorLast{Golm {et~al.}} 
\def\Authors{Jessica Golm\,$^{1,2,*}$, Jose Mar\'ia ~Garc\'ia-Barcel\'o\,$^{3,*}$, Sergio Arguedas Cuendis\,$^{1}$, Sergio Calatroni\,$^{1}$, Walter Wuensch\,$^{1}$, and Babette D\"obrich\,$^{3}$ }
\def\Address{$^{1}$CERN - European Organization for Nuclear Research, Geneva, Switzerland\\
$^{2}$Friedrich  Schiller  University  Jena (Institute  for  Optics  and  Quantum  Electronics), Jena, Germany\\
$^{3}$Max-Planck-Institut f\"{u}r Physik (Werner-Heisenberg-Institut), Boltzmannstr. 8, 85748 Garching bei M\"{u}nchen, Germany }

\def\corrAuthor{Jessica Golm, Jose Mar\'ia Garc\'ia-Barcel\'o\ }

\def\corrEmail{jessica.golm@cern.ch, jmgarcia@mpp.mpg.de}

\begin{document}
\onecolumn
\firstpage{1}

\title {Split-cavity tuning of a rectangular axion haloscope operating around 8.4 GHz} 

\author[\firstAuthorLast ]{\Authors} 
\address{} 
\correspondance{} 

\extraAuth{}

\maketitle

\begin{abstract}
The axion haloscope is the currently most sensitive method to probe the vanishingly small coupling of this prominent Dark Matter candidate to photons. To scan a sizeable axion Dark Matter parameter space, the cavities that make up the haloscope need to be tuned efficiently. In this article, we describe a novel technique to tune axion haloscopes around $8.4$~GHz in a purely mechanical manner without the use of dielectrics. We achieve tuning by introducing a gap along the cavity geometry. A quality factor reduction of less than 20\% is achieved experimentally for a tuning range of around 600~MHz at room temperature and at cryogenic temperatures for around 300~MHz. A larger tuning range would require an improved alignments mechanism. We present the results of a corresponding prototype and outline prospects to further develop this technique.
\end{abstract}

\section{Introduction}  
\label{sec:TUNING}
\graphicspath{{Tuning_pics/}}

An axion haloscope is an experimental apparatus used in the search for axions, hypothetical particles that could make up Dark Matter in the Universe as discussed in \cite{Peccei:1977Jun,Peccei:1977Sep}. In recent years, there has been a large increase in the number of experiments and approaches that aim to find or exclude the axion as Dark Matter, see e.g. here \cite{Irastorza:2018dyq}.

Axions are very light, neutral particles that are expected to interact very weakly with ordinary matter and electromagnetic fields. Axion haloscopes are designed to detect axions by converting them into detectable electromagnetic signals (photons) in the presence of a strong magnetic field by the inverse Primakoff effect, cf. \cite{Primakoff:1951}. To improve this conversion process, haloscopes use high-quality factor resonators, such as microwave cavities, see e.g. here \cite{Sikivie:1983ip}. The conversion of axions into photons occurs when the mass of the axion matches approximately the resonant frequency of the apparatus.

The primary goals of an effective axion detection system involve maximising the power generated from the axion-photon interaction and expanding the range of analysed axion masses (i.e. the range of frequencies scanned). The detected radio frequency (RF) power ($P_d$) relies on inherent axion properties and experimental cavity characteristics, as outlined in \cite{RADESreviewUniverse}:

\begin{equation}
\label{eq:Pd}
    P_d \, = \, \kappa \, g^2_{a\gamma}\, \frac{\rho_a}{m_a} \, B_e^2 \, C \, V \, Q_l,
\end{equation}

Here, $\kappa$ signifies the coupling to the external receiver ($\kappa=0.5$ for critical coupling operation), $g_{a\gamma}$ represents the unknown axion-photon coupling, $\rho_a$ denotes Dark Matter density, $m_a$ stands for axion mass, $B_e$ represents the external static magnetic field (dependent on the magnet used in the experiment), $C$ is a label for the so-called `form factor', $V$ is the cavity volume, and $Q_l$ represents the loaded quality factor of the cavity. In contrast to $Q_l$, the unloaded quality factor $Q_0$ is a common metric in characterising resonant cavities as it is independent of external coupling, see e.g. here \cite{pozar,Rezaee:2012}. The form factor ($C$), measuring the coupling between the external magnetostatic field and the RF electric field induced by axion-photon conversion, is expressed as:

\begin{equation}
\label{eq:C}
    C \, = \, \frac{|\int _V \, \vec{E} \cdot \vec{B}_e \, dV|^2}{\int_V \, |\vec{B}_e|^2 \, dV \int_V \, \varepsilon_r \, |\vec{E}|^2 \, dV},
\end{equation}

Here, $\varepsilon_r$ represents the relative electrical permittivity within the cavity of volume $V$. The sensitivity measure of the haloscope is the axion-photon coupling detectable for a given signal-to-noise ratio ($\frac{S}{N}$), which can be calculated using:

\begin{equation}
\label{eq:ga}
    g_{a\gamma} \, = \,  \left(\frac{\frac{S}{N} \, k_B \, T_{sys}}{\kappa \, \rho_a \, C \, V \, Q_l}\right)^{\frac{1}{2}}\frac{1}{B_e}\left(\frac{m_a^3}{Q_a \, \Delta t}\right)^{\frac{1}{4}},
\end{equation}

In this equation, $k_B$ represents the Boltzmann constant, $T_{sys}$ is the noise temperature of the system, $Q_a$ represents the quality factor of the axion resonance, and $\Delta t$ is the data-taking time window, cf. \cite{RADESreviewUniverse}. To summarise, the controllable parameters in a given experimental set-up encompass $\kappa$, $C$, $V$, and $Q_l$.

On the other hand, the mass of axions, if they exist, is not known. To maximise the chances of detecting axions, the haloscope needs to be tuned to the resonant frequency that corresponds to the expected mass range of axions being searched for. Tuning ensures that the cavity is effectively only sensitive to the specific axion masses under investigation, improving the chances of successful detection. Thus, the scanning rate $\frac{dm_a}{dt}$ is commonly employed to assess the performance of a haloscope, a value derived from Equation \ref{eq:ga}, cf. \cite{Kim_2020}:

\begin{equation}
\label{eq:dmadt}
    \frac{dm_a}{dt} \, = \, Q_a \, Q_l \, \kappa^2 \, g_{a\gamma}^4 \, \left(\frac{\rho_a}{m_a} \right)^2 \, B_e^4 \, C^2 \, V^2 \, \left(\frac{S}{N} \, k_B \, T_{sys}\right)^{-2}.
\end{equation}
Also, $\kappa$ is usually rewritten in terms of the coupling coefficient $\beta$, which can be extracted from the expression
\begin{equation}
\label{eq:betakappa}
    \beta = \frac{\kappa}{1-\kappa}.
\end{equation}

A tuning step must not require the opening of the cryostat as this is very time and energy-consuming. Until today, most of the adjustment techniques applied in haloscope experiments for frequency tuning have relied on mechanical systems. Various collaborations such as ADMX (cf. \cite{Boutan:2018} and \cite{PhysRevLett.130.071002}), and HAYSTAC (cf. \cite{Zhong:2018}), utilise cylindrical cavities with metallic rods. Also, other teams like IBS/CAPP (cf. \cite{Choi:2021}) employ dielectric rods. These rods are rotated inside the cavities, altering the electromagnetic field distribution and, consequently, the resonant frequency of the operating mode. This rotation is achieved either through several gears connected to a motor or by employing piezoelectric elements. A slightly different technique was, for example, employed by the QUAX collaboration (cf. \cite{Alesini:2020}), by using adjustable sapphire shells to modify the resonant frequency of the cavity. In a different example, the CAST-CAPP/IBS group uses two movable dielectric sapphire plates positioned symmetrically at the cavity sides (see e.g. here \cite{Miceli:2015}). In \cite{Kuo_2021} tunable conic shell-cavities have been investigated, providing good tuning results for large cavities operating around $7.5$ and $20$~GHz. Table~\ref{tab:tuningrefs} gives an examination of some current tuning systems for frequencies between $4$ and $11$~GHz, in comparison with the outcomes obtained in this work.
\begin{table}[h]
\scalebox{0.8}{
\begin{tabular}{|c|c|c|c|c|c|}
\hline
Experiment & Freq. (GHz) & Tuning ($\%$) & $Q_0/10^4$ & $C$ & References \\ \hline\hline
CAST-CAPP & $4.774-5.174$ & $7.7$ & $\sim4$ & $\sim0.53$ & \cite{Adair:2022rtw} \\ \hline
ADMX Sidecar - Run B & $5.086-5.799$ & $12.3$ & $0.22\times(1+\beta)$ & $0.44-0.61$ & \cite{Boutan:2018} \\ \hline
HAYSTAC & $5.6-5.8$ & $3.45$ & $1.8$ & $\sim0.5$ & \cite{Zhong:2018} \\ \hline
ADMX Sidecar - Run C & $7.173-7.203$ & $0.42$ & $0.23\times(1+\beta)$ & $0.040-0.046$ & \cite{Boutan:2018} \\ \hline
RADES - Vertical Cut & $7.753-8.420$ & $9.52$ & $ 1.5-2.7$ & $\sim0.66$ & this work \footnotemark   \\ \hline
QUAX & $10.1-10.3$ & $2$ & $\sim9$ & $\sim0.11/V$ & \cite{10.1063/5.0137621} \\ \hline
\end{tabular}
}
\centering
\caption{\label{tab:tuningrefs} Overview of some existing tunable haloscopes operating at frequencies between $4$ and $11$~GHz. For the last row, the form factor is given in terms of volume ($C/V$), where $V$ is in L. Also, for the second and fourth rows, the unloaded quality factor parameter has been left dependent on the $\beta$ coupling employed for each corresponding experiment.}
\end{table}
\footnotetext{The tuning range presented in the table for this work was achieved experimentally at cryogenic temperatures as well as the quality factor (the upper Q-value was obtained with spacers, the lower value was the minimum measured while tuning with a gear system). The form factor presented was taken from simulations for parallel cavity halves. }
The percentage value of tuning has been calculated according to the following equation
\begin{equation}
\label{eq:Tuning}
    \mathrm{Tuning} = \frac{f_2-f_1}{f_2}\times100 \quad [\%],
\end{equation}
where $f_2$ is the upper limit in the frequency tuning range and $f_1$ is the lowest frequency. It is worth mentioning that the ADMX Sidecar experiment has conducted axion data taking campaigns using two different resonance modes. For Run B, this group has employed the $TM_{010}$ mode, while for Run C they have used $TM_{020}$, to explore higher frequencies, where significantly lower form factor values are achieved. On the other hand, the CADEx collaboration proposed in \cite{Aja_2022} the use of a sliding or tunable wall mechanism in rectangular cavities. The ORGAN group, in \cite{McAllister:2023ipr}, also describes this type of mechanical tuning in detail, comparing it with the tuning technique using metal rods usually employed in the axion community. In addition, the QUAX group has explored in \cite{10.1063/5.0137621} opening techniques in a cylindrical cavity using angular misalignment on the longitudinal axis, achieving satisfactory results at frequencies above $10$~GHz, using a so-called `clamshell' cavity.
A tunable rectangular axion cavity is described in \cite{McAllister:2023ipr}. It was used in The ORGAN experiment Phase 1b, where the axion search was focused on the 26–27 GHz band. A movable side wall (moving plunger) provided the tuning. 

It should be mentioned that there is also the possibility of broad-band search, which removes the need of tuning, but typically cannot achieve a signal amplification comparable to the resonant haloscopes. Examples of such experiments are BRASS (magnetic mirror), see \cite{Bajjali:2023uis}, MADMAX (with a set of movable dielectric disks in \cite{MADMAX:2019pub}), ALPHA (using metamaterials), cf.\cite{PhysRevD.102.043003}, or BREAD, see \cite{Knirck:2023jpu}.

Over the past five years, the RADES (Relic Axion Dark-Matter Exploratory Setup) research group has extensively explored haloscopes, engaging in the conceptualisation, production, and physics analysis of axion detectors in the pursuit of detecting Dark Matter axions with masses approximately around $1$ and $34 \, \mu$eV , cf. \cite{RADES_paper1,RADES_paper2,RADES_paper3,RADESreviewUniverse,Ahyoune:2023gfw}. This team has developed a mechanical tuning technique where the haloscope volume is increased by mechanically moving the cavity halves split symmetrically (cf. \cite{RadesProceeding:2020}), operating at X-band frequencies. This set-up can work similarly to the clamshell technique (cf. \cite{10.1063/5.0137621}), but a symmetric opening is more favourable, giving a bigger frequency tuning range, as shown below. Also, the RADES group is currently exploring other methods based on electrical tuning by the use of ferroelectric materials (see e.g. here \cite{Garcia:2023}).

In this work, our goal is to build a tuning mechanism that can tune the resonant frequency of the first RADES cavity (cf. \cite{RADES_paper1}) by using the tuning technique presented in \cite{RadesProceeding:2020} with a custom made support structure which gives the capability of tuning and quality factor measurements at cryogenic temperatures.
This halo scope is based on an array of five subcavities interconnected by waveguide irises to increase the total volume while maintaining a high resonant frequency for operation in dipole magnets. The frequency tuning approach presented in this study relies on the separation of two symmetrical halves until RF leakage through the aperture becomes the primary factor affecting $Q_0$. In the case of small apertures in this haloscope, the effect of a vertical cut along the longitudinal direction is negligible, as the microwave currents of the fundamental mode $TE_{101}$ are parallel to this plane. This movement effectively increases the width of the cavity, thereby altering the search range for axion frequencies. According to \cite{RadesProceeding:2020}, the theoretical tuning range value for this haloscope is around $10$~$\%$. Similar values are also obtained in \cite{Volume_paper} for this type of tuning at X-band frequencies in other cavities. For this study, experiments are carried out at cryogenic temperatures, and simulations using the finite element method are employed to comprehend how the cavity behaves at various openings.

\section{Cavity design for mechanical tuning}

\begin{figure} [htb]
    \begin{minipage}{\columnwidth}
        \centering
        \includegraphics[width=0.6\textwidth]{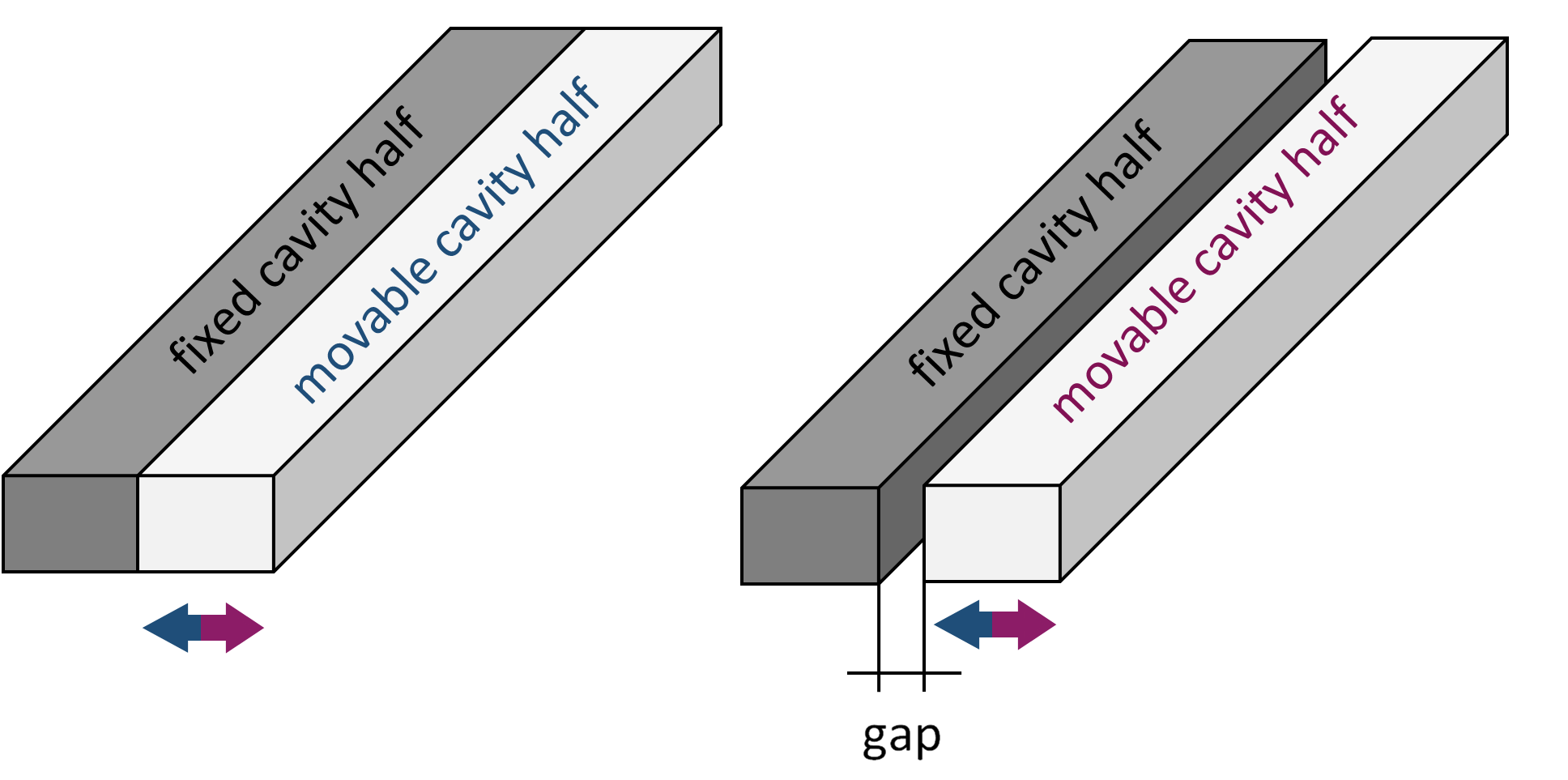}
    \end{minipage}
    \caption{Schematics of the tuning concept. The left side shows the cavity in the closed position with no gap and at its highest resonant frequency. On the right side, a gap was introduced between the two cavity halves, tuning the cavity to a lower frequency depending on the gap size.}
    \label{fig:schematicsTuning}
\end{figure}

Fig.~\ref{fig:schematicsTuning} shows the schematics of the mechanical tuning idea. By introducing a gap between two cavity halves the geometry can be widened which shifts the resonant frequency to lower frequencies\footnote{In \cite{Volume_paper}, preliminary investigations have been conducted with this technique for rectangular long single cavities, and 1D and 2D multicavity structures operating at X-band frequencies. A similar research is expected for the long single cavities, and 1D and 2D haloscopes with cylindrical shapes explored in \cite{Cyl_paper}.}.

\subsection{Cavity design}

The tuning mechanism will be exemplified using multi-cell cavities, which were developed by the RADES team, as described in \cite{RADES_paper1}. This design allows an increase in cavity volume while maintaining a high resonant frequency as a single element of the cavity mostly determines the frequency. A prototype resonating at $8.4$~GHz was built by joining five cells with all-inductive irises (vertical windows in waveguide technology). While the tuning will be demonstrated with this prototype, it works in principle for other designs as well.

\begin{figure} [htb]
	\begin{minipage}{\columnwidth}
		\centering
		\includegraphics[width=0.7\textwidth]{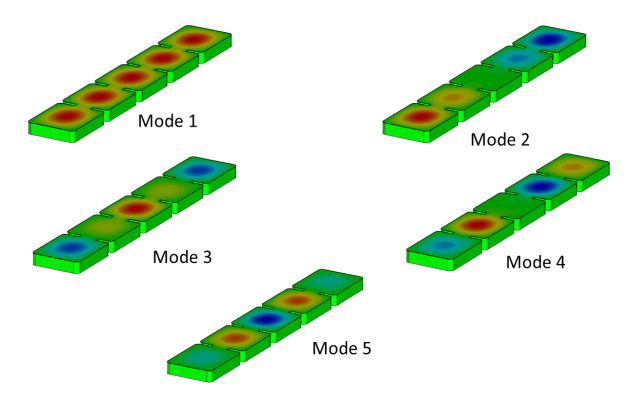}
	\end{minipage}
    \caption{Electric field pattern for the five electromagnetic modes of the cavity. Red colour refers to a maximum and blue colour to a minimum in the electric field. Taken from \cite{RADES_paper1}.}
    \label{fig:Fieldpattern}
\end{figure}

\begin{figure}[htb]
    \begin{minipage}[t]{0.33\linewidth}
        \includegraphics[width=1\textwidth]{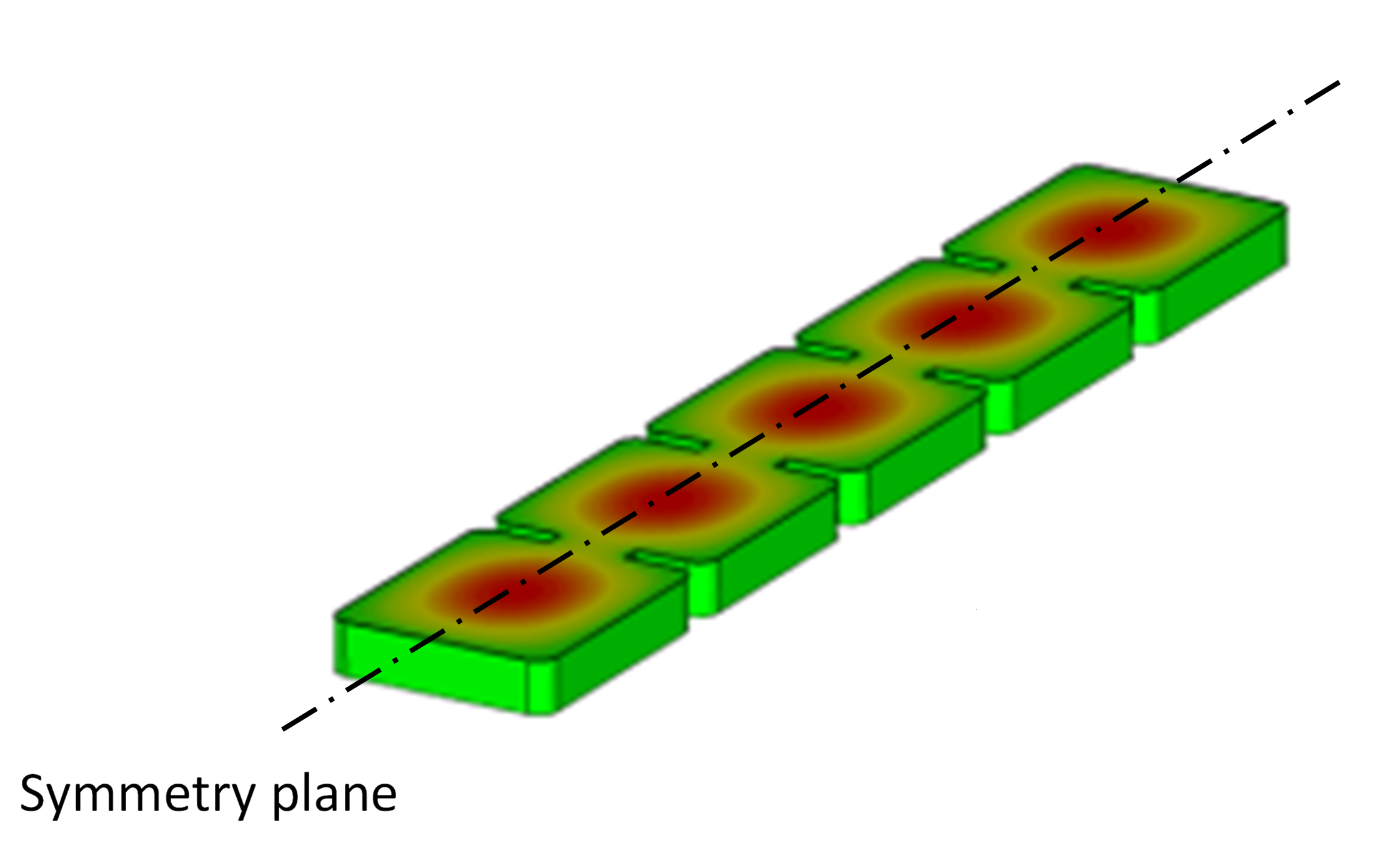}	
    \end{minipage}
    \hfill
    \begin{minipage}[t]{0.6\linewidth}
        \centering
        \includegraphics[width=1\textwidth]{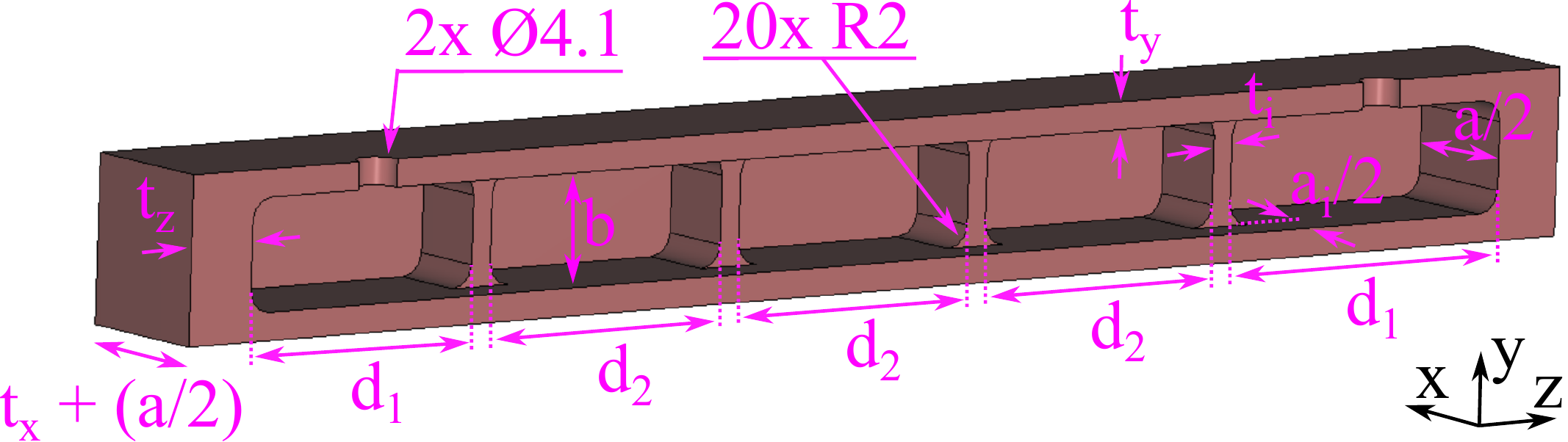}
    \end{minipage}
    \caption{Left: Mode pattern with symmetry plane. The fabrication is done in two halves defined by this symmetry plane. 3D model housing of the vertical cut inductive irises cavity to be manufactured (right). This piece is one of the two symmetrical halves, which must be parallel. SMA coaxial ports are situated at the $\varnothing4.1$ holes. Dimensions given in Table~\ref{tab:VC_dimensions}.}
    \label{fig:5Iris_vs_tuning}
\end{figure}

Fig.~\ref{fig:Fieldpattern} shows the electric field distribution of the five configuration modes.
The coherence between the cavities is maintained only in the first mode which is the only mode that couples to the axion. To tune this type of cavity, it was cut along the symmetry plane as indicated in Fig.~\ref{fig:5Iris_vs_tuning} (left).
The surface currents do not cross this plane of symmetry (they are parallel to this plane); therefore, it is possible to cut the cavity without intercepting the current. Thus, one can expect to retain a high $Q_0$ value despite the introduction of the manufacturing cut and even when separating both halves. This paper aims to study and measure to what extent this is possible.

To enable vertical cut tuning, small changes in the design of the 5-cell cavity (with manufacturing horizontal cut), described in \cite{RADES_paper1}, were necessary. This cut can be used to introduce a gap that increases the volume and moves the resonant modes to lower frequencies, as shown in Fig.~\ref{fig:schematicsTuning}, without a significant loss of quality factor, as will be shown in the following. In general terms, the gap should be small enough with $gap<\lambda$/2, where $\lambda$ is the wavelength at the operation frequency, see e.g. here \cite{Torrisi2020}. However, in our case, as a more conservative scenario, a gap has been set to be small compared to the cavity dimensions ($<10$~$\%$ of the width, a value extracted from \cite{Volume_paper}, where a similar tuning mechanism is employed). Fig.~\ref{fig:5Iris_vs_tuning} (right)\footnote{In the fabrication of these parts, alignment pins were included in the $x-$axis, in the thickness $t_z$ of the housing at both ends in length. In addition, screw holes were implemented both in the port sections ($y-$axis) and in the $t_z$-thickness ($x-$axis). The latter to test the behaviour of the completely closed and screwed-together structure.} depicts the 3D model and dimensions of the analysed haloscope. In Table~\ref{tab:VC_dimensions}, the dimension values are shown.
\begin{table}[h]
\begin{tabular}{|c|c|}
\hline
Dimensions & Values (in mm) \\ \hline\hline
$a$ & $22.86$ \\ \hline
$b$ & $10.16$ \\ \hline
$d_1$ & $26.68$ \\ \hline
$d_2$ & $25$ \\ \hline
$a_i$ & $8$ \\ \hline
$t_i$ & $2$ \\ \hline
$t_x$ & $2.5$ \\ \hline
$t_y$ & $2.5$ \\ \hline
$t_z$ & $6.5$ \\ \hline
\end{tabular}
\centering
\caption{\label{tab:VC_dimensions} Dimension values of the vertical cut haloscope for cryogenic conditions analysed in this study.}
\end{table}

\subsection{Simulations of losses introduced due to mechanical tuning}
\label{ssec:Simulation}
\subsubsection{Parallel cavity halves}
\label{ssec:Simulation_parallel_cavities}

Before the manufacturing and measurements of this haloscope, a study was carried out using the \cite{CST_and_FEST3D}, to study the behaviour of this structure in terms of $Q_0$ and C factors against different openings that can cause radiation losses.

\begin{figure}[htb]
    \begin{minipage}[t]{0.33\linewidth}
        \includegraphics[width=1\textwidth]{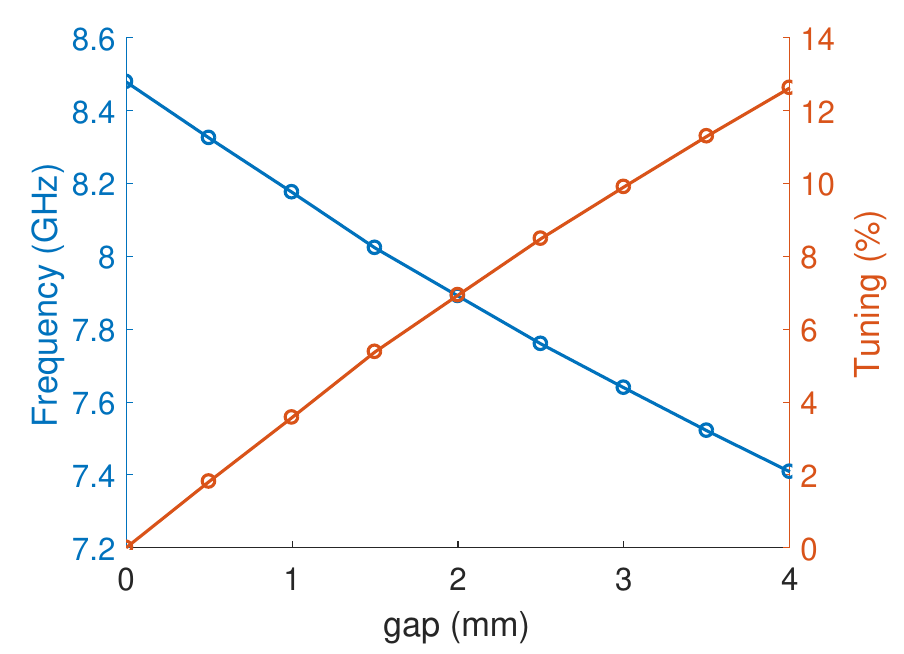}	
    \end{minipage}
    \begin{minipage}[t]{0.33\linewidth}
        \centering
        \includegraphics[width=1\textwidth]{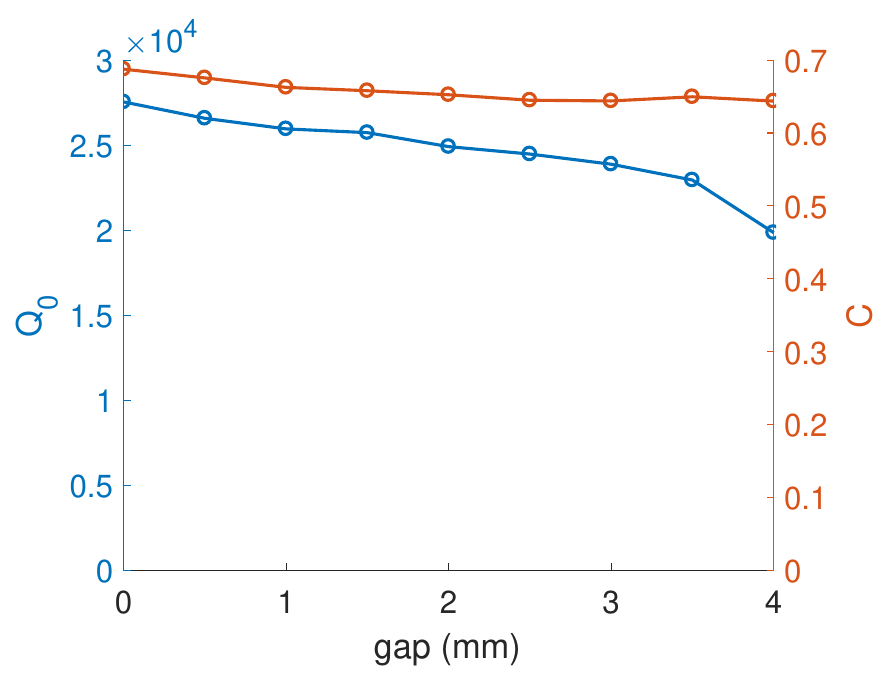}
    \end{minipage}
    \begin{minipage}[t]{0.33\linewidth}
        \centering
        \includegraphics[width=1\textwidth]{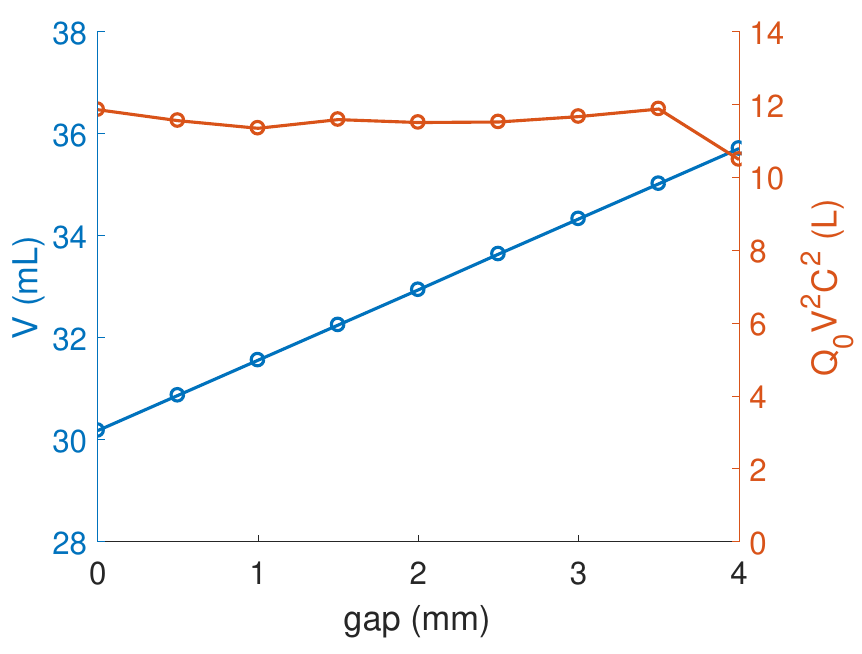}
    \end{minipage}
    \caption{Results from CST simulations for the vertical cut tuning study applied in the haloscope depicted in Figure~\ref{fig:5Iris_vs_tuning} (right): frequency and tuning versus gap (left), unloaded quality and form factors versus gap (centre), and volume and figure of merit ($Q_0V^2C^2$) versus gap (right).}    \label{fig:VC_verticalcuttunigresults_CST}
\end{figure}

Fig.~\ref{fig:VC_verticalcuttunigresults_CST} depicts the results obtained with a sweep of vertical cut gaps from $0$ to $4$~mm.
For these simulations, an electrical conductivity value in the copper cavity of $\sigma_c=10^9$~S/m was used, as this is a realistic value according to \cite{9699394} where the same coating and frequency were applied. In addition, coaxial antenna lengths have been chosen to be critically coupled ($\beta=1$) for Port~$1$ and very undercoupled ($\beta\simeq0$) for Port~$2$, in the closed scenario (gap$=0$~mm). The antenna lengths have been maintained for the study of all the gap values.

As can be seen, the obtained form factor follows a relatively constant value, close to $C=0.66$, for the whole range. However, the unloaded quality factor suffers a small detriment when the gap increases. For the full range of gaps analysed, the figure of merit $Q_0V^2C^2$ remains almost constant at around $11.5$~L, so it can be guaranteed that the tuning range is adequate, obtaining a frequency tuning range of $1.07$~GHz (or $12.6$~$\%$), which is a very satisfactory tuning compared to the results of other experimental groups shown in Table~\ref{tab:tuningrefs}, and even more so for operating frequencies close to $8.4$~GHz.

In addition, since this structure has a tuning system that connects to the outside environment, a study has been made of possible signals that could enter its interior. After several simulations, it has been verified that there are no resonance echoes of a potential dipole magnet with a bore diameter of about $50$~mm where this haloscope could be installed.

\subsubsection{Misalignment studies}\label{ssec:MisalingmentSim}

\begin{figure}[htb]\centering
    \begin{minipage}[t]{0.69\linewidth}
        \centering
        \includegraphics[width=1\textwidth]{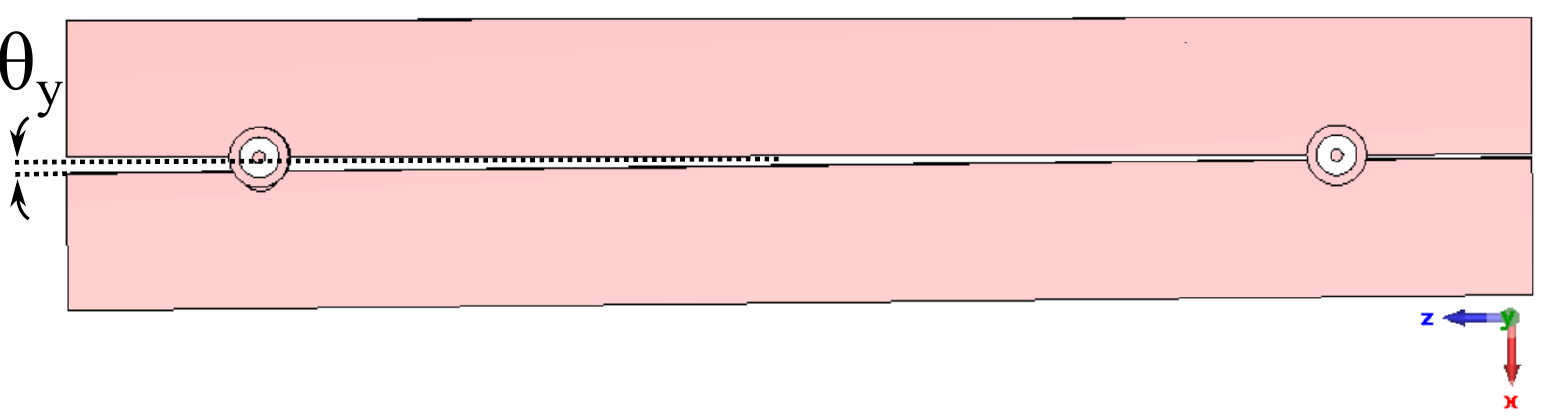}
    \end{minipage}
    \begin{minipage}[t]{0.25\linewidth}
        \centering
        \includegraphics[width=1\textwidth]{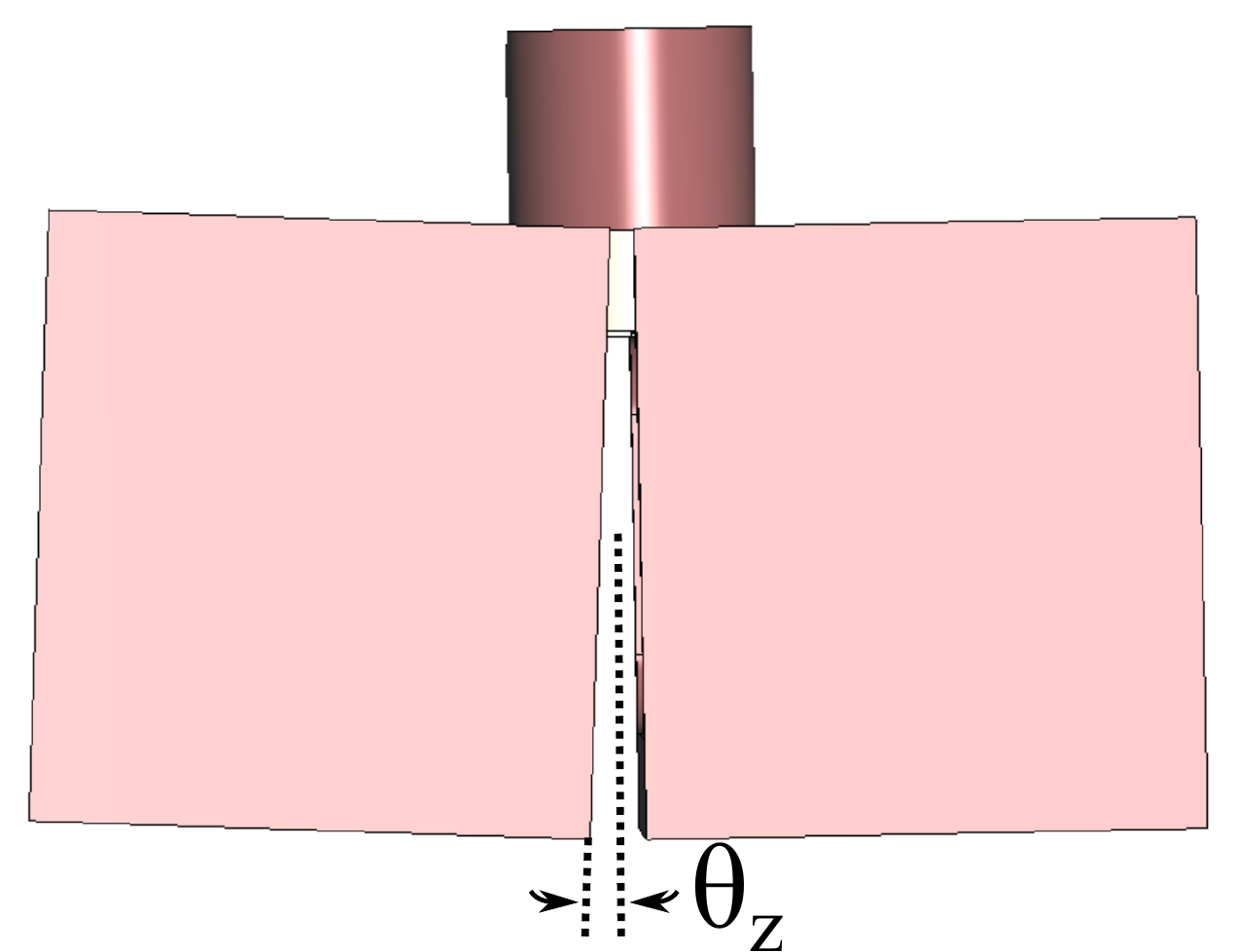}
    \end{minipage}
    \begin{minipage}[t]{0.45\linewidth}
        \centering
        \includegraphics[width=1\textwidth]{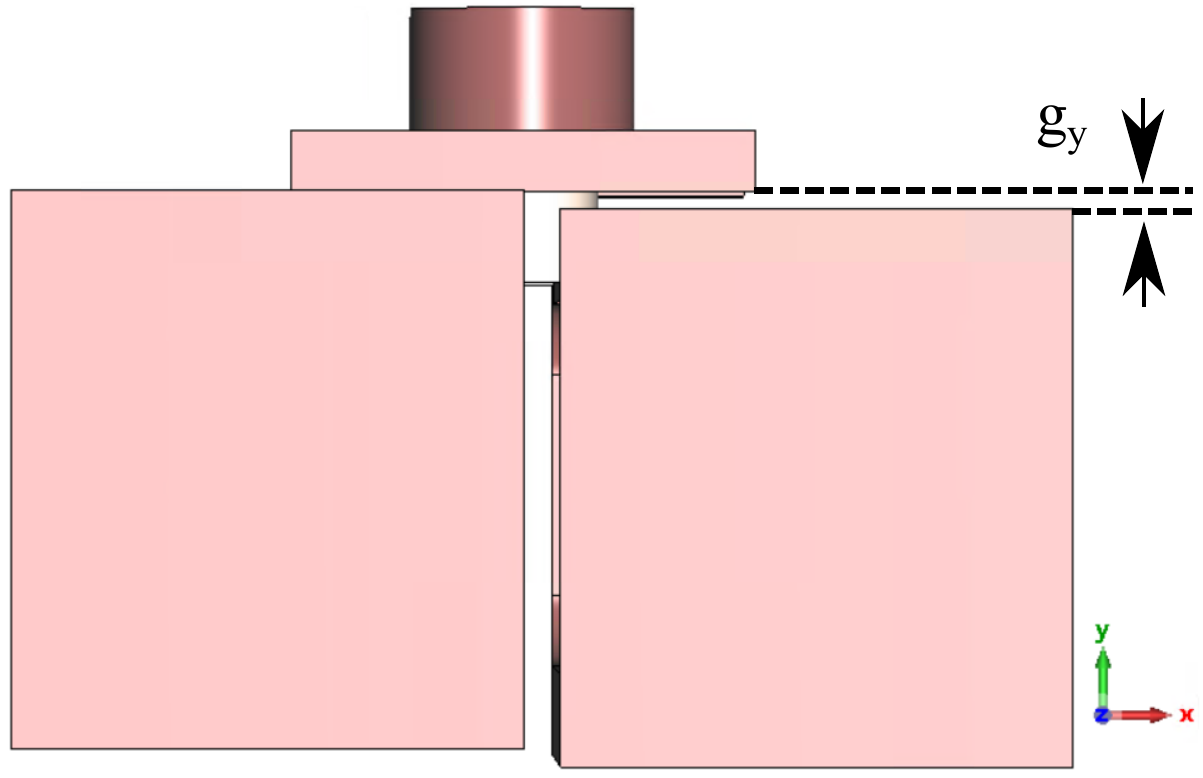}
    \end{minipage}
    \caption{Misalignment effect in the vertical cut haloscope for three scenarios: angular $y-$axis (top left), angular $z-$axis (top right), and lineal $y-$axis (bottom).}
    \label{fig:VC_Tilt}
\end{figure}

A simulation study has been conducted to consider possible misalignment effects when separating the two halves. Due to the position of the coaxial ports, the number of possible main misalignments is reduced to the following: angular $y-$axis, angular $z-$axis, and lineal $y-$axis. Fig.~\ref{fig:VC_Tilt} shows an example of misalignment for each scenario.
In addition, to reduce complexity, this tuning technique is based on leaving the coaxial ports attached to one of the halves of the haloscope.  This introduces a small asymmetry between the two halves which can be neglected (see Appendix A). 

\begin{figure}[htb]
    \begin{minipage}[t]{0.33\linewidth}
        \centering
        \includegraphics[width=1\textwidth]{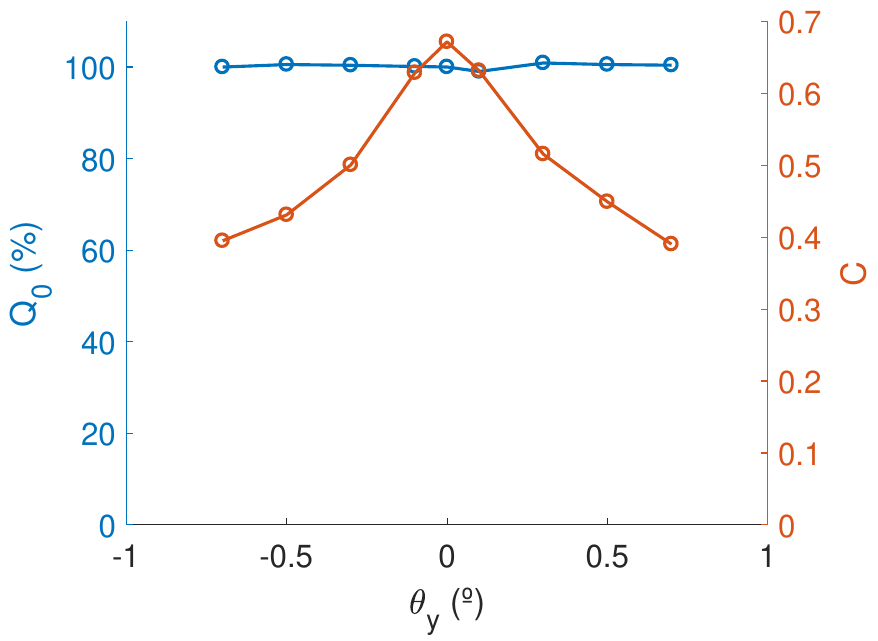}	
    \end{minipage}
    \begin{minipage}[t]{0.33\linewidth}
        \centering
        \includegraphics[width=1\textwidth]{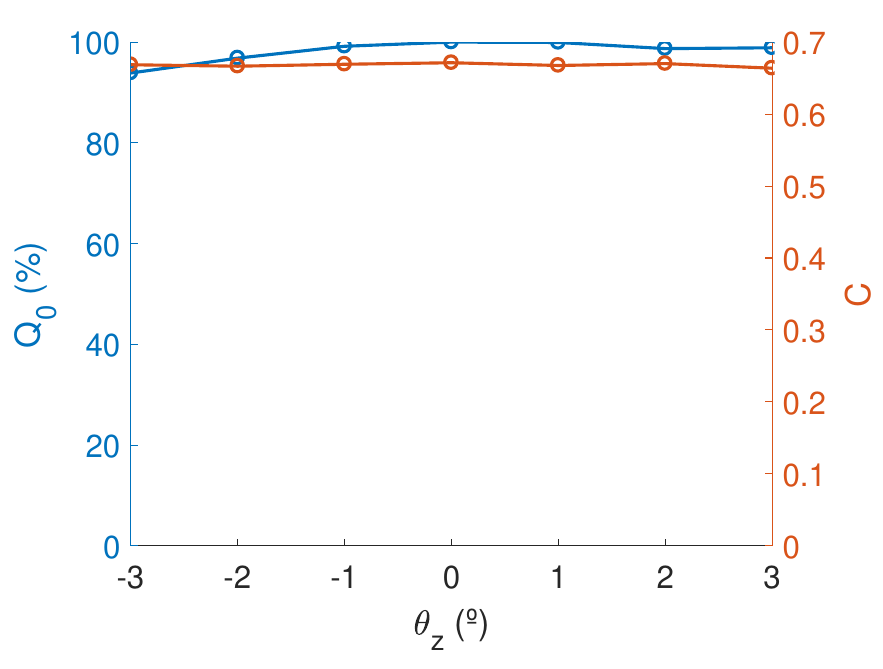}	
    \end{minipage}
    \begin{minipage}[t]{0.33\linewidth}
        \centering
        \includegraphics[width=1\textwidth]{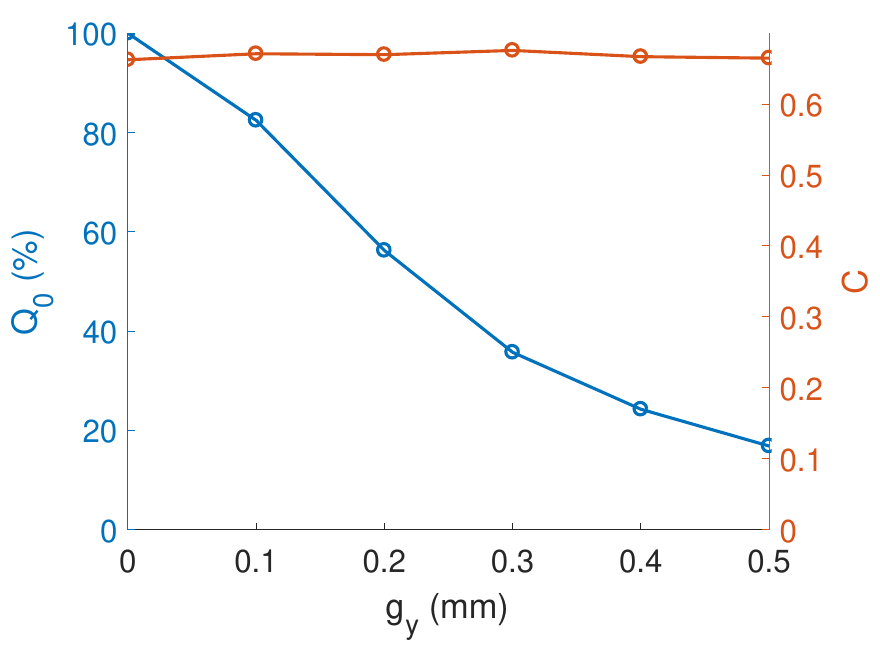}	
    \end{minipage}
    \begin{minipage}[t]{0.33\linewidth}
        \centering
        \includegraphics[width=1\textwidth]{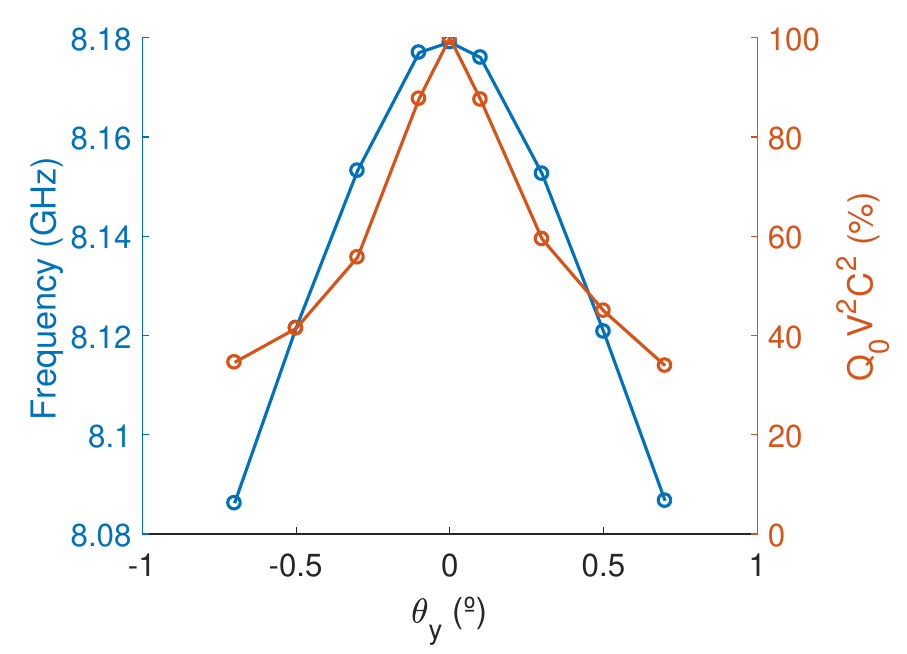}
    \end{minipage}
    \begin{minipage}[t]{0.33\linewidth}
        \centering
        \includegraphics[width=1\textwidth]{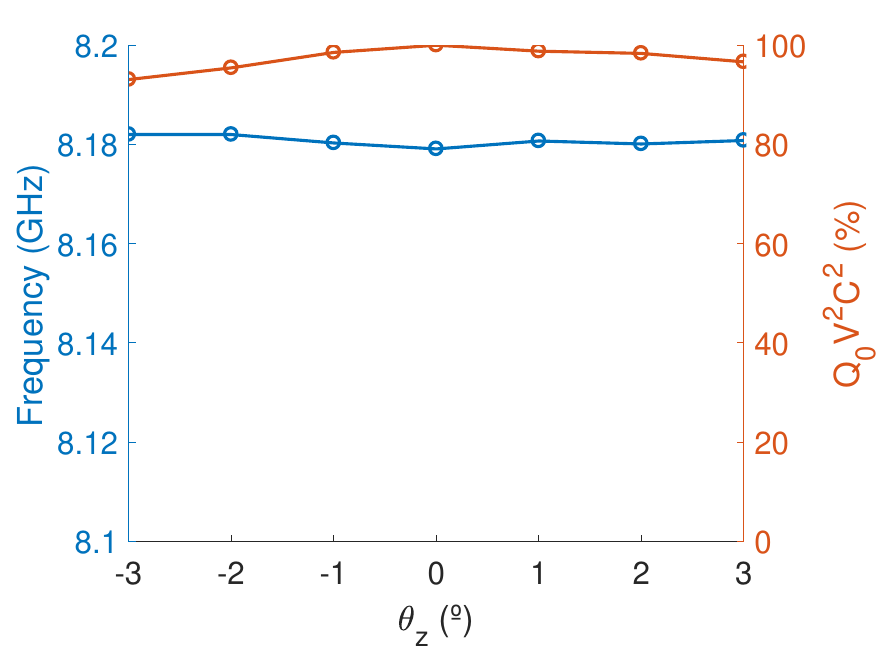}
    \end{minipage}
    \begin{minipage}[t]{0.33\linewidth}
        \centering
        \includegraphics[width=1\textwidth]{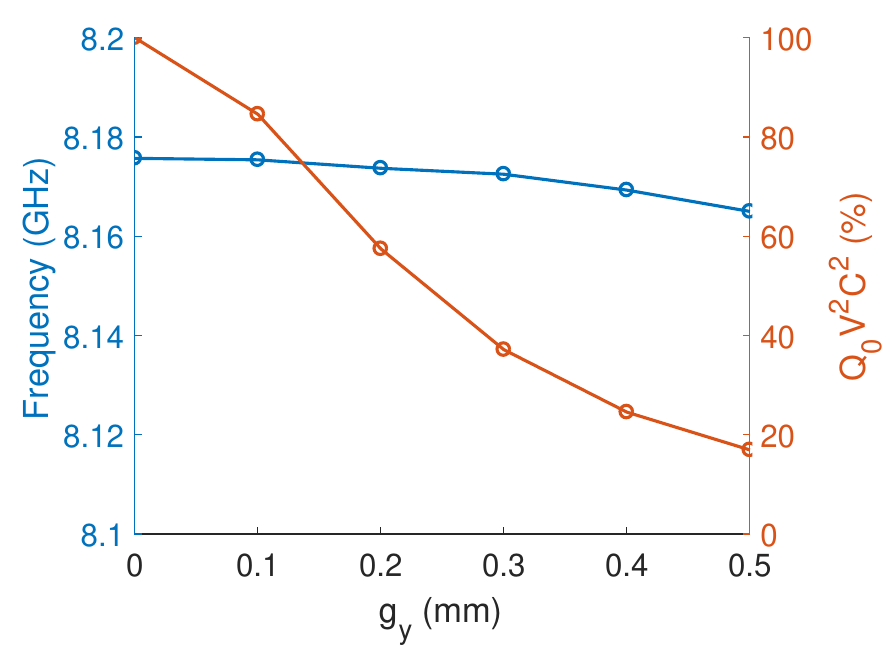}
    \end{minipage}
    \caption{Results obtained in the simulation of the misalignment study in the angular $y-$axis (first column), in the angular $z-$axis (second column), and in the lineal $y-$axis (third column). The first row of graphs shows the variation of the quality and form factors versus the variation of the misalignment variable. The second row plots the variation in frequency and figure of merit $Q_0V^2C^2$ versus misalignment. The quality factor and figure of merit parameters are given in terms of the percentage change from the aligned scenarios ($\theta_y = \theta_z = g_y = 0$).}
    \label{fig:VC_Misalignment_CST}
\end{figure}

Also, to simplify this investigation of misalignments, a gap of $1$~mm has been chosen. Fig.~\ref{fig:VC_Misalignment_CST} depicts the results obtained for each case as a function of the misalignment variables: $\theta_y$ for the angular $y-$axis, $\theta_z$ for the angular $z-$axis, and $g_y$ for the linear $y-$axis.
As can be seen, the results show a degradation in the form factor value of $\sim40$~$\%$ in the case of angular misalignment in the $y-$axis. The main reason for this reduction is because the individual resonant frequencies of the subcavities are very sensitive to their widths, which are modified with $\theta_y$, in turn changing the optimised condition for the operating mode (see more on this concept in \cite{RADES_paper1}). Furthermore, for this type of misalignment, the resonant frequency changes considerably (around $1.2$~$\%$). However, the unloaded quality factor remains relatively stable.

On the other hand, for angular $z-$axis misalignment, the form factor remains very high, although the quality factor suffers a reduction of more than $7$~$\%$ for negative values of $\theta_z$. However, as the form factor becomes more important here for the figure of merit of the experiment, the latter remains relatively high. In the case of frequency, it is also maintained around stable values.

Finally, in the case of linear misalignment in the $y-$axis, it is seen that this causes the largest decrease of the unloaded quality factor (more than $80$~$\%$ for values of $g_y = 0.5$~mm), while the form factor remains high for the entire range, and this in turn reduces the figure of merit the most of the three cases analysed. Fortunately, the outcomes show a resonant frequency that is relatively stable for the entire range.

\section{Proof of principle}
\label{sec:PROOFofPri}

The tuning idea was first implemented by separating the cavity halves of a prototype with spacers that do not penetrate the cavity hollow volume or disturb the electromagnetic field pattern. The prototype was made of non-magnetic stainless steel (316 LN) coated with a $30$~$\mu$m copper layer. The reason for choosing stainless steel as a base material for the cavity and not producing it from pure copper is the forces the cavity can be exposed to during a quench of the magnet (eddy currents). Cavities for axion Dark Matter searches are usually installed in multitesla fields.

\begin{figure} [htb]
	\begin{minipage}{\columnwidth}
		\centering
		\includegraphics[width=0.5\textwidth]{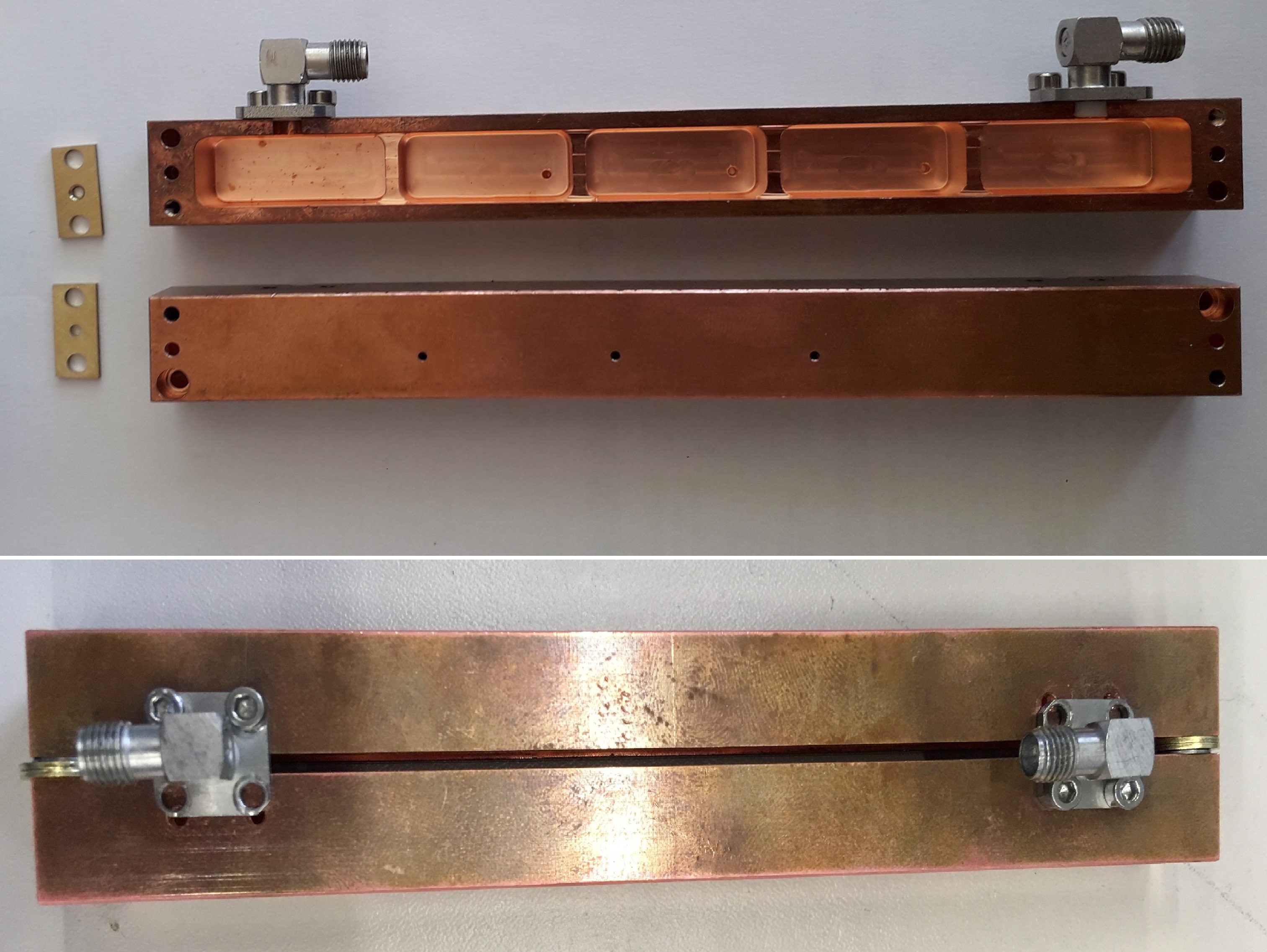}
	\end{minipage}
    \caption{Photograph of the vertical cut haloscope coated with copper before assembly and brass spacers (top) and after assembly with a gap between the two halves introduced with stainless steel washer spacer (bottom). Note that for the characterisation both ports were attached to one cavity half unlike shown on the picture in which each port is attached to another cavity half.}
    \label{fig:outside_tuning}
\end{figure}

A photograph of the prototype and spacers is shown in Fig.~\ref{fig:outside_tuning}. The top picture shows the disassembled cavity halves and the spacers made of brass. These spacers generate a gap of around $1.25$~mm and they are mounted on the screws that connect the two cavity halves. The bottom picture depicts the assembled cavity with spacers made of stainless steel. After a characterisation of the tuning properties of the cavity at room temperature, the cavity was installed in a cryostat, and the unloaded quality factor $Q_0$ was measured below 10 K for a gap size of $0$ and $1.25$ mm. The brass spacers were used during the cryogenic characterisation. In none of the described mounting situations, the spacers entered the cavity structure. The alignment of the two halves of the cavity with each other depended only on the two spacers and the force with which the screws were tightened on each side. Both cavity ports were kept on one cavity half when a gap was applied.

\begin{figure}[htb]
	\begin{minipage}[t]{0.5\linewidth}
		\centering 
		\includegraphics[width=1\textwidth]{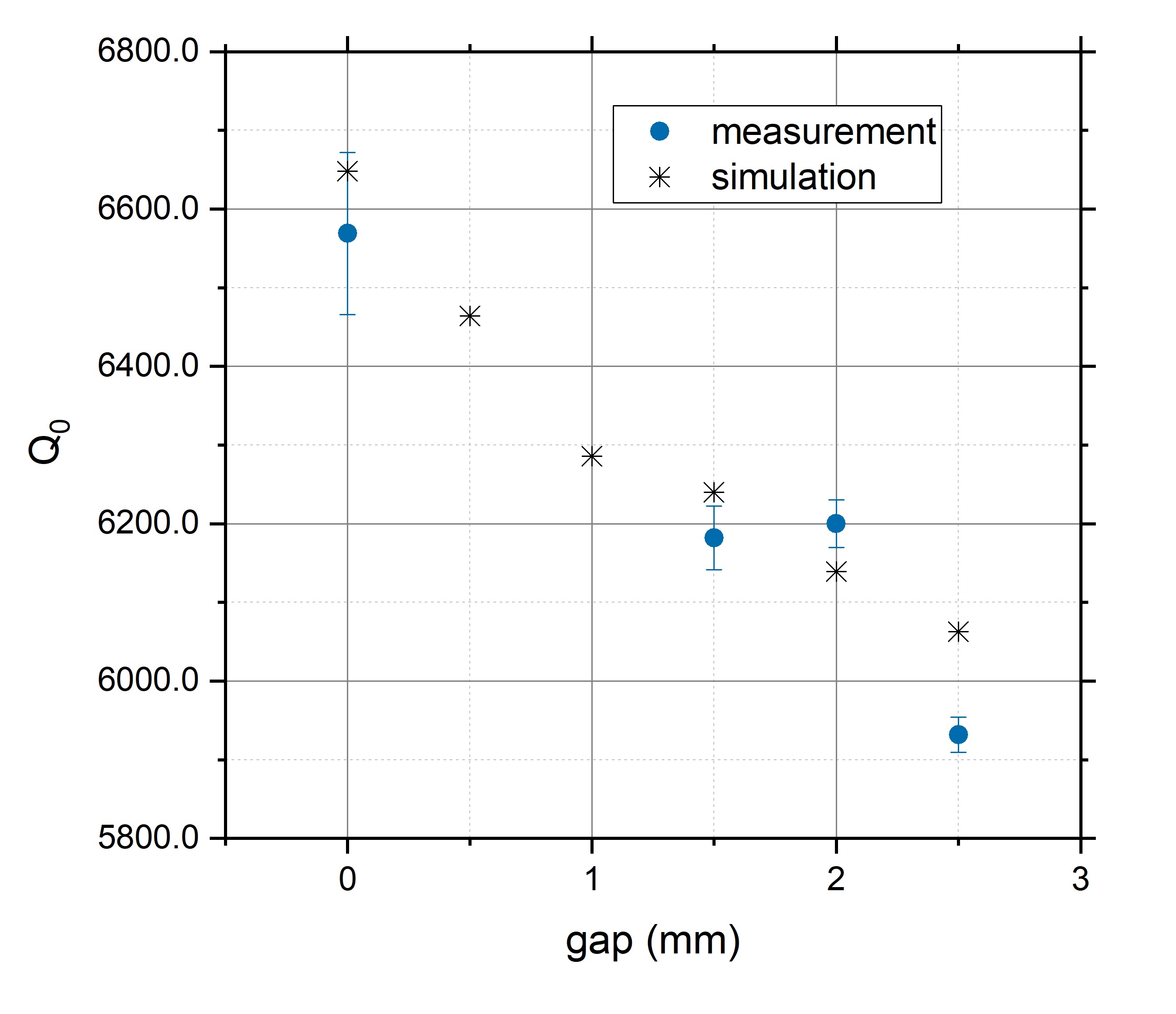}
	\end{minipage}
    \hfill
	\begin{minipage}[t]{0.5\linewidth}
		\centering 
		\includegraphics[width=0.96\textwidth]{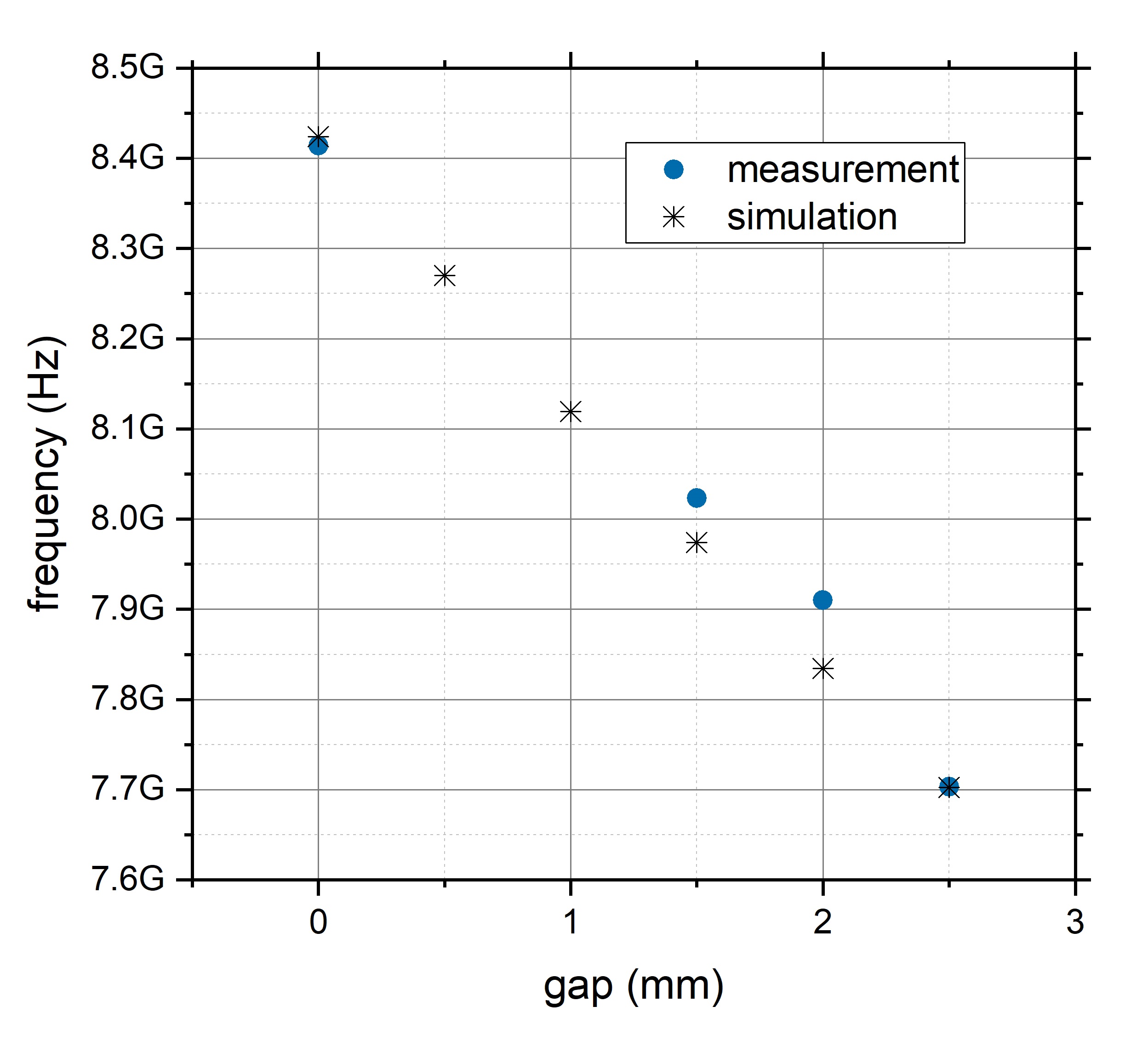}
	\end{minipage}
	\caption{Comparison of measurement with uncertainty (blue) and simulation (black) of the unloaded quality factor $Q_0$ (left) and the resonant frequency (right) for different gap openings. For the simulations a copper conductivity at room temperature of $5.8\times10^7$~$S/m$ was assumed.}
	\label{fig:CavityProp_tuning2}
\end{figure}

Fig.~\ref{fig:CavityProp_tuning2} shows measurements of the tuning range and quality factor at ambient conditions and compares it with simulation. For this characterisation of the cavity, washers of known thickness were used to create a gap while a vice pressed the cavity halves together. There is a linear dependence between gap size and frequency, as expected, and the total tuning range at ambient conditions is $711$~MHz for a gap opening of $2.5$~mm, as shown in Fig. ~\ref{fig:CavityProp_tuning2} (right). There is only a small deviation between measurement and simulation in the unloaded quality factor of  Fig.~\ref{fig:CavityProp_tuning2} (left), but by differently assembling the cavity halves losses can increase. From simulations in section~\ref{ssec:MisalingmentSim} we know these losses can be attributed to misalignment.

\begin{table}[htb]
\centering
\begin{tabular}{|c|c|c|c|}
\hline
gap (mm) & frequency (GHz)  & unloaded quality factor Q$_0$     & electrical conductivity (S/m) \\ \hline
0       & 8.446634 $\pm$ 0.000004 & 28300 $\pm$ 600 &  $101\times10^7$    \\ \hline
1.25    & 8.111632 $\pm$ 0.000020 & 27200 $\pm$ 1500 &   $103\times10^7$ \\ \hline
\end{tabular}
\caption{Unloaded quality factor $Q_0$ and resonant frequency $f$ from a cavity measurement and uncertainty for temperatures below $10$~K and gap sizes of $0$ and $1.25$~mm. An electrical conductivity was deduced from these measurements.}
\label{tab:cryo_measurements}
\end{table}

For the characterisation at low temperature, the cavity was installed in a vacuum chamber which was introduced in a helium cryostat. The cavity was suspended in the cryostat by clamping it in a support structure on one extremity and subsequently attached to a long rod. It was measured in two configurations during cool-down for a gap size of $0$ and $1.25$~mm (half the maximum possible opening within the holding structure mechanics) separated by the brass spacers shown in Fig.~\ref{fig:outside_tuning} (top). The results from the cryogenic measurements of the quality factor with spacers below 10K are shown in Tab.~\ref{tab:cryo_measurements}. From those measurements, the electrical conductivity of the copper coating can be deduced assuming that the pieces are aligned. These conductivity values are also shown in Tab. ~\ref{tab:cryo_measurements} and agree well with those measured from previous coatings on other samples ($100\times10^7$~$S/m$).
The copper coating of this cavity was applied by DC (Direct Current) galvanic plating with an organic brightener and the RRR (residual-resistance ratio) is expected to be $38$~$\pm$~$2$.

Misalignment of the cavity halves can result in high losses. This was observed when remeasuring the cavity after reassembly. The unloaded quality factor obtained at low temperatures was measured to $Q_0$~(gap$=0$~mm)~$=$~$23200$~$\pm$~$800$ and $Q_0$~(gap $=1.25$~mm)~$=$~$20500$~$\pm$~$1100$. The quality factor decreased by more than 20~\% which is consistent for example with a misalignment in direction g$_y$ of about 100 $\mu$m according to section~\ref{ssec:MisalingmentSim}.

Due to the form factor studies in simulation considering misalignments of the cavity halves, we can justify that this parameter will move around the ranges shown in Fig.~\ref{fig:VC_Misalignment_CST}. Thus, an argument for explaining the real form factor value arises. Two measurements at cryogenic temperatures
gave $\Delta f_0$ (difference in the resonant frequency between both measurements) of less than $0.03$~$\%$ for both gaps $0$ and $1.25$~mm (see Table~\ref{tab:cryo_measurements}). Compared to the misalignment simulations this low $\Delta f_0$ gives us that we have a very low misalignment scenario with $\theta_y$ (see Fig.~\ref{fig:VC_Misalignment_CST}), so the form factor is not affected by this. Also, due to the high $\Delta Q_0$ (difference in the unloaded quality factor between both measurements) ($20$~$\%$) we can affirm that we have linear misalignments $g_y$ (probably around $0.1$~mm), but the form factor is neither affected by this misalignment. Thus, observing Fig.~\ref{fig:VC_Misalignment_CST} and considering the above information the C-factor should be around $0.62$.

\section{Tuning mechanism for cryogenic environment }
\subsection{Set-up and sliding mechanism }
\label{ssec:sliding mechanism}

After demonstrating that the tuning concept is viable in principle, a gear system was designed to allow for translating the movement in a cryogenic environment. For data-taking, haloscope axion cavities will be installed in dipole magnets where the space for tuning is typically limited by small bore diameters. Furthermore, the measurement takes place in a vacuum or liquid helium environment at temperatures of $4.2$~K.
\begin{figure} [htb]
    \centering 
    \includegraphics[width=0.7\textwidth]{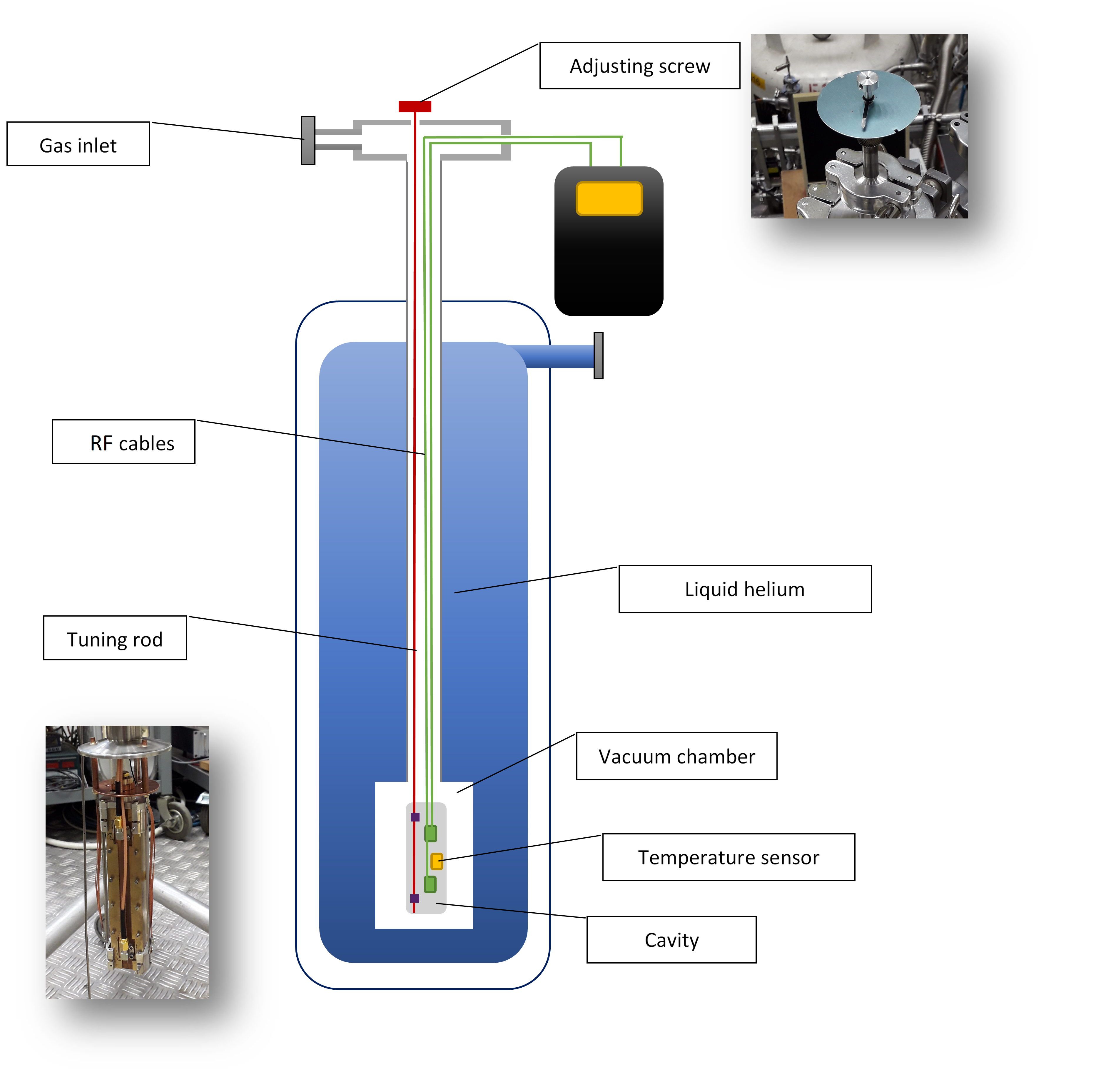}
    \caption{Cavity tuning test stand in the CERN Central Cryogenic laboratory with adjusting screw and rod (red), RF cables (green), temperature sensor (yellow), and the cavity (grey).} 
    \label{fig:CLSetup}
\end{figure}
These conditions make it challenging to realise cavity tuning during data-taking. Therefore, two possible solutions were deliberated. One option is to install a piezo actuator in the bore to generate the movement. However, two challenges have to be overcome: first, the space in the bore does not allow for a larger actuator. Secondly, an actuator with sufficient force to move the cavity halves is needed. Considering the size of the currently existing actuators that can move our cavity halves, a gear system to generate the movement would be necessary. Another possibility is to generate the movement from the outside of the magnet by a long rod connected via gears to the two cavity halves. This solution is also not without problems as complex feedthroughs are required to propagate the rotational movement from the $300$~K to the $4$~K environment.

To study the gear-based tuning mechanism, a test stand was built in the CERN Central Cryogenic Laboratory, and different aspects of this tuning were investigated. Fig.~\ref{fig:CLSetup} shows a schematic drawing of the test stand. The cavity and the RF cables were installed in a vacuum chamber immersed in liquid helium. To characterise the cavity two RF cables (green) coming from the cavity were connected to a feedthrough in the vacuum flange at ambient temperature, to which a Vector Network Analyser (Keysight N9918A) was connected for quality factor measurements. To tune and change the gap size of the cavity a long rod (about $1.5$~m long, $2$~mm diameter, stainless steel) (red) was connected by a gear system to the cavity at the bottom of the vacuum chamber. The rod had an adjusting screw at the end and by turning it the cavity opened or closed, depending on the direction in which the gear is turned. The movement of the rod for this test was generated by hand, but the adjusting screw could be easily attached to a small motor which would allow more precision in movements. A gas inlet on the cryostat provided the possibility of measuring in a helium gas environment instead of in vacuum.

\begin{figure}[]
    \begin{minipage}{\columnwidth}
    \centering 
    \includegraphics[width=0.99\textwidth]{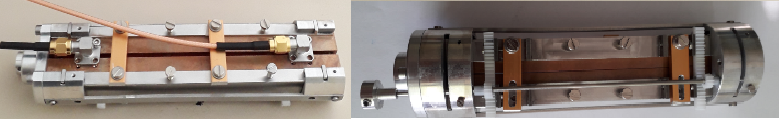}
    \end{minipage}
    \caption{Photographs of the cavity halves installed in the tuning holding structure with gears. Top view with RF ports (left) and bottom view (right). The bottom view displays the gear mechanism that moves the cavity halves to each other.}
    \label{fig:photoAssembl1}
\end{figure}

\begin{figure} [htb]
    \centering 
    \includegraphics[width=0.99\textwidth]{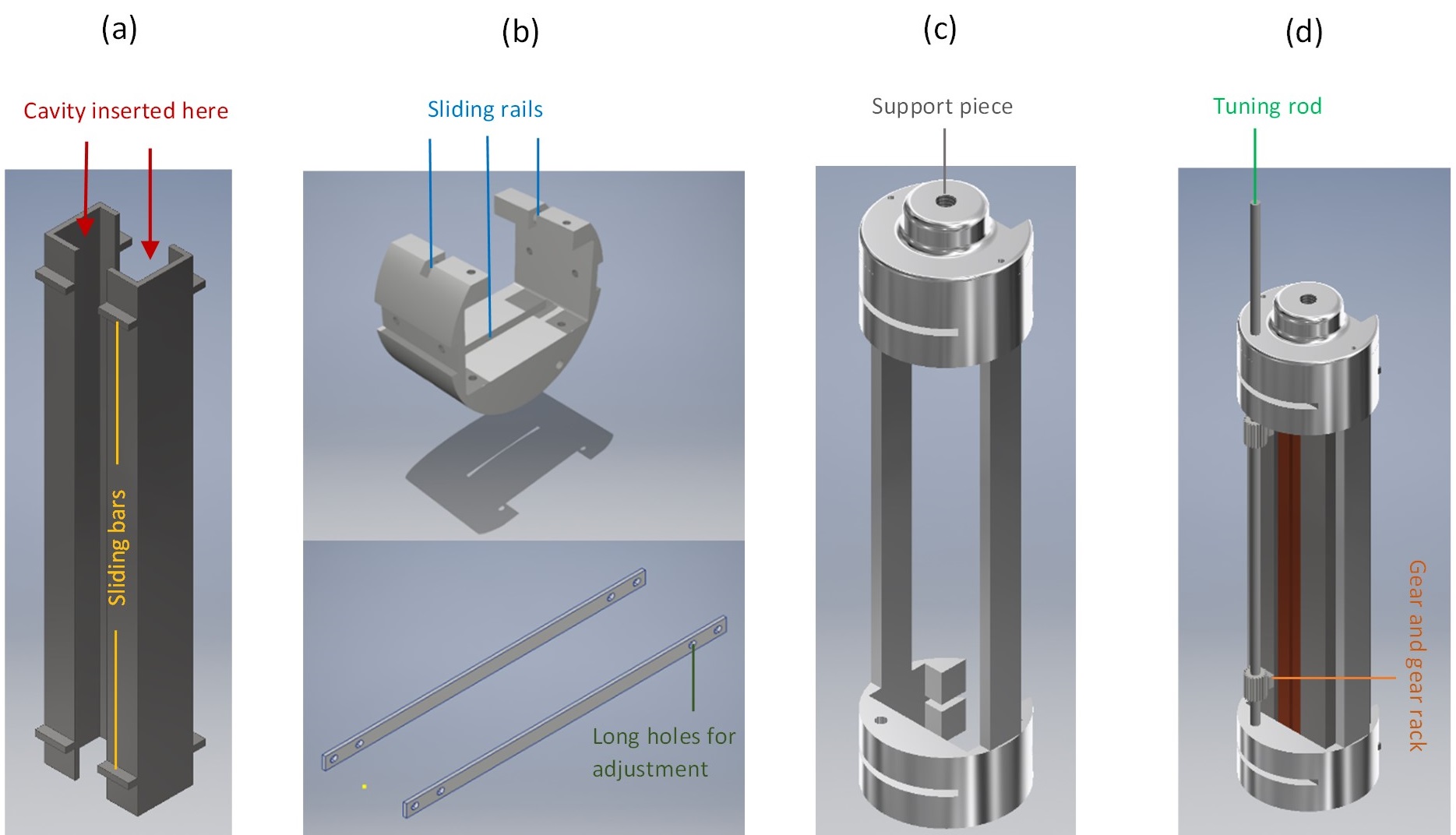}
    \caption{Drawings of the cavity support structure with (a) holding piece for the cavity halves, (b) the sliding structure for alignment, (c) the assembly of the sliding structure, and (d) the complete assembly with the gear system.} 
    \label{fig:Tuning5c_assemblyDRW}
\end{figure}

Photographs of the cavity installed in the holding system with gears to transmit the movement can be seen in Fig.~\ref{fig:photoAssembl1}.
In Fig.~\ref{fig:Tuning5c_assemblyDRW}, the individual parts of the tuning mechanism are shown. In (a), the holding pieces for the cavities are illustrated. The cavity halves can slide into these holding pieces, which are held by positioning screws. The holding piece has on both sides sliding bars. In Fig.~\ref{fig:Tuning5c_assemblyDRW} (b), the sliding structure is shown. The assembly consists of two of those structures that will be connected by long stainless steel bars, which have on one side long holes for easier positioning. The sliding bars shown in (a) have to be inserted in the sliding rails shown in (b). Fig.~\ref{fig:Tuning5c_assemblyDRW} (c) shows the assembly of (a) and (b) without cavities, and (d) shows the complete assembly with the cavity halves installed. The assembly shows the two gear racks that are screwed to holding piece (a) and a tuning rod with two gears glued to it. Thus, by turning this tuning rod outside of the cryostat, the rotation will be transmitted to the gears on the cavity inside the cryostat and converted into a linear, parallel translation of one cavity half. 

To guarantee a smooth movement of the cavity halves with the described tuning mechanism at cryogenic temperatures and in vacuum, the materials have to be chosen carefully. In vacuum, parts of the same material can easily adhere to each other because there are no thin layers of other elements between the parts. Under mechanical pressure, e.g. thermal contraction, they can bond. Also, the friction coefficients between the same materials, i.e. aluminium on aluminium (friction coefficient $1.6-2.2$, c.f. \cite{friction}) or stainless steel on stainless steel (friction coefficient $2.9$, see \cite{friction}), in vacuum are very high. Thus, it was decided to produce the holding pieces for the cavity and the holding structure out of different materials. The holding piece for the cavity was made of stainless steel (see Fig.~\ref{fig:Tuning5c_assemblyDRW} (a)) and the structure on which the cavity slides back and forth inside was made of aluminium (friction coefficient $0.3$, see \cite{friction}, and Fig.~\ref{fig:Tuning5c_assemblyDRW} (b)). The gear and gear rack were made of PTFE (Polytetrafluoroethylene) which has a very low friction coefficient of $0.04$ for PTFE on PTFE, cf. \cite{friction}.

After a first functionality test of the described mechanism and material, some optimisations have been made. The sliding structure (see Fig.~\ref{fig:Tuning5c_assemblyDRW} (a)) was re-fabricated using brass as it has a more similar expansion coefficient to aluminium than stainless steel. Furthermore, the holding pieces for the cavity were optimised by replacing the sliding bars with sliding pins, which have less surface area touching the sliding rail, resulting in reduced friction and less material deforming during cool-down.

\subsection{Tuning results with the sliding mechanism}

\begin{figure} []
	\begin{minipage}[t]{0.49\linewidth}
		\centering 
		\includegraphics[width=1\textwidth]{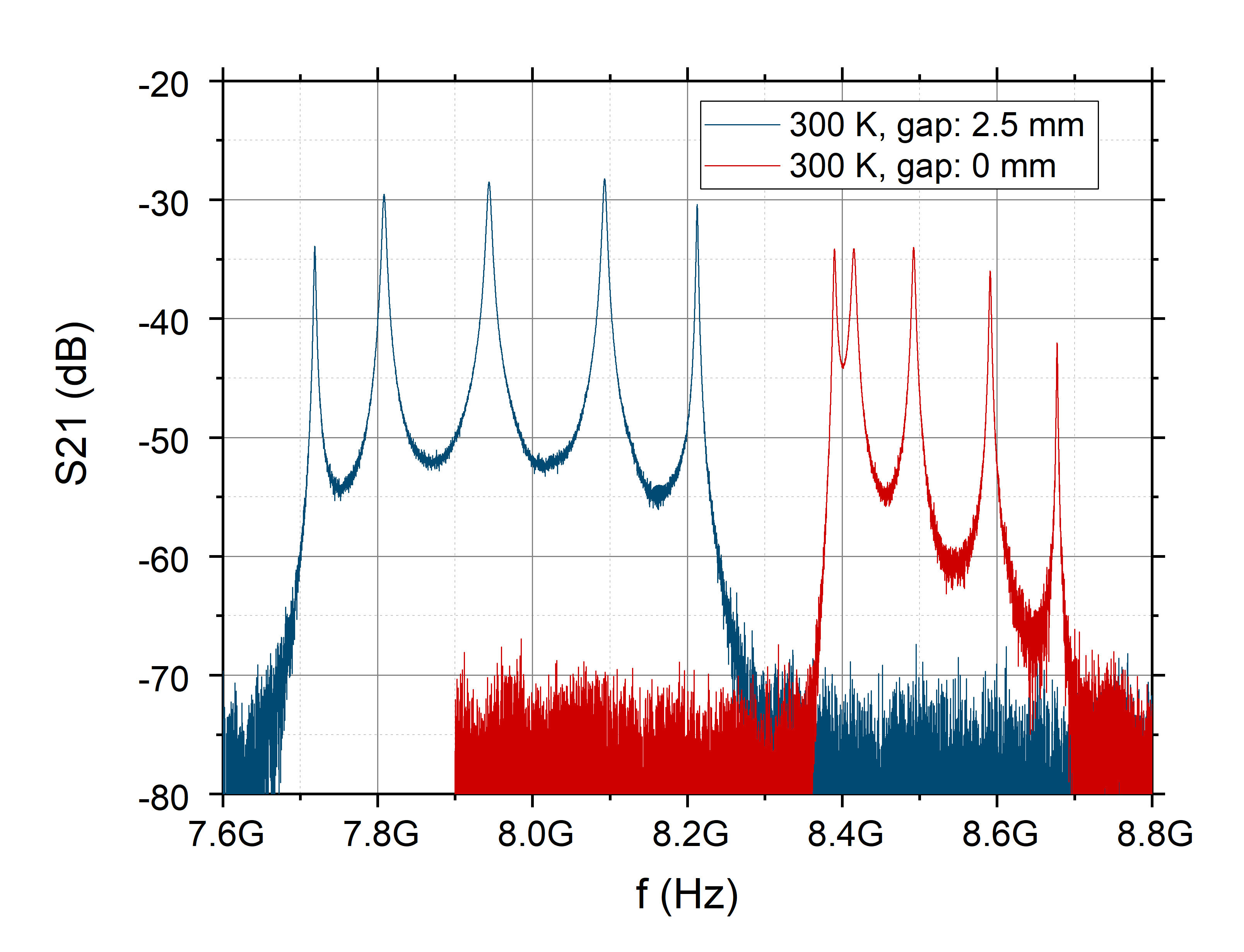}
	\end{minipage}
    \hfill
	\begin{minipage}[t]{0.49\linewidth}
		\centering 
		\includegraphics[width=01\textwidth]{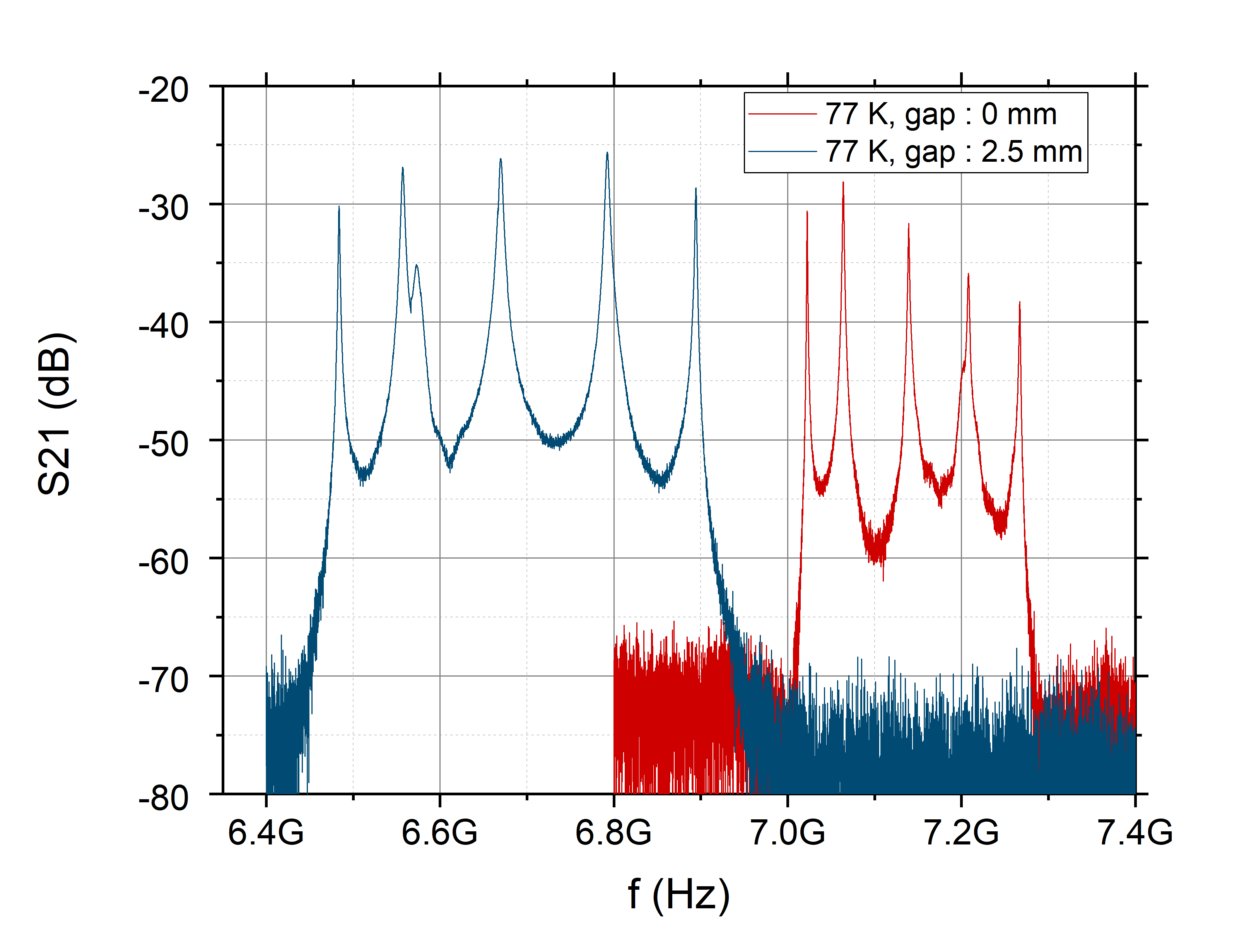}
	\end{minipage}
	\caption{Tuning range measured with the cavity embedded in the sliding structure at ambient conditions (left) and in liquid nitrogen (right) with a maximum gap size of $2.5$~mm (blue) and no gap (red). At $77$~K disturbances of the spectra due to the boiling of liquid nitrogen are visible, most clearly for the second cavity peak in the spectra of the cavity with a $2.5$~mm gap.}
	\label{fig:AIRvsLN2}
\end{figure}

The tuning mechanism described in the previous section was tested on functionality at $77$~K by immersing the set-up in liquid nitrogen. For testing a tuning mechanism where pieces are made of materials with different thermal expansion coefficients, it is crucial to wait for thermalisation. After a 5-minute thermalisation time in the liquid bath, an opening and closing of the cavity halves were demonstrated. The cavity spectrum and the tuning range were measured at ambient conditions, see Fig.~\ref{fig:AIRvsLN2} (left), and in liquid nitrogen, see Fig.~\ref{fig:AIRvsLN2} (right). The spectrum of the cavity shows five peaks, one for each of the small subcavities, while the axion only couples to the first mode. For explanations, see \cite{RADES_paper1}. The difference in frequency between the first peak of each spectrum (and thus the tuning range) in the open (gap$ = 2.5$~mm) and closed (gap$ = 0$~mm) positions of the cavity halves was measured to be $670$~MHz at ambient condition and $540$~MHz in liquid nitrogen at $77$~K. The tuning range in liquid nitrogen is smaller compared to air because the higher dielectric constant of $\textrm{LN}_2$ decreases the resonant frequency by a factor of $1/\sqrt{\varepsilon}$. The dielectric constant of liquid nitrogen is $1.431$, see \cite{LNdielectric:1967}, and therefore a tuning range of $540$~MHz between $6.48$ and $7.02$~GHz corresponds to a tuning range in vacuum of about $650$~MHz between $7.75$ and $8.40$~GHz.

In Fig.~\ref{fig:AIRvsLN2} (right) some disturbances in the second peak of the blue spectrum are visible. Those have their origin in gaseous Nitrogen bubbles that enter the cavity from time to time. Those bubbles are the result of the boiling of the liquid nitrogen and change the spectra while entering the cavity.

After demonstrating that the mechanism worked in a $77$~K environment, the whole set-up was implemented in a cryostat with a $2$~mm thick tuning rod of about $1.5$~m length going from $4$ to $300$~K. The cavity and all elements were placed in a vacuum chamber inside the cryostat that was filled with liquid helium. A schematic of the experimental set-up is shown in Fig.~\ref{fig:CLSetup}. It was possible to reach temperatures around $20$~K in the vacuum chamber with this setup. The reason for not reaching lower temperatures is the RF cables which are made of copper, a very good thermal conductor, as well as the tuning rod of $1.5$~m length (stainless steel, $2$~mm diameter) conducting heat from the outside of the cryostat at ambient temperature to the cold vacuum chamber. For measurements with spacers between the cavity halves, described in section~\ref{sec:PROOFofPri}, a minimum temperature of $6$~K could be reached. However, this was only possible due to a complex thermalisation of the RF cables to intercept the conducted heat into the vacuum chamber. Those cables need to be moveable during tuning so as not to obstruct the mechanism,  therefore they were not thermalised. Instead, thinner, more flexible cables were used with the disadvantage of higher RF losses.

The same results in terms of tuning are expected at $4$ and $20$~K, as previous measurements showed that the resonant frequency (= thermal contraction of the cavity) and quality factor (= electrical conductivity) for the cavity are constant below $30$~K, see Fig.~\ref{fig:CavityProp_tuning2} (right). To use the cavity as a haloscope in an axion search, the temperature should be as low as possible. One possibility to solve the issue of not reaching the minimum temperature is to let a small amount of exchange gas (about $1$~mbar of helium) in the vacuum chamber, allowing convective heat transfer from the wall of the vacuum chamber to the cavity. By adding the exchange gas, the cavity properties remain the same, and the gas will not affect the haloscope data taking.

 \begin{figure}[h]
	\begin{minipage}[t]{0.49\linewidth}
		\centering 
		\includegraphics[width=1\textwidth]{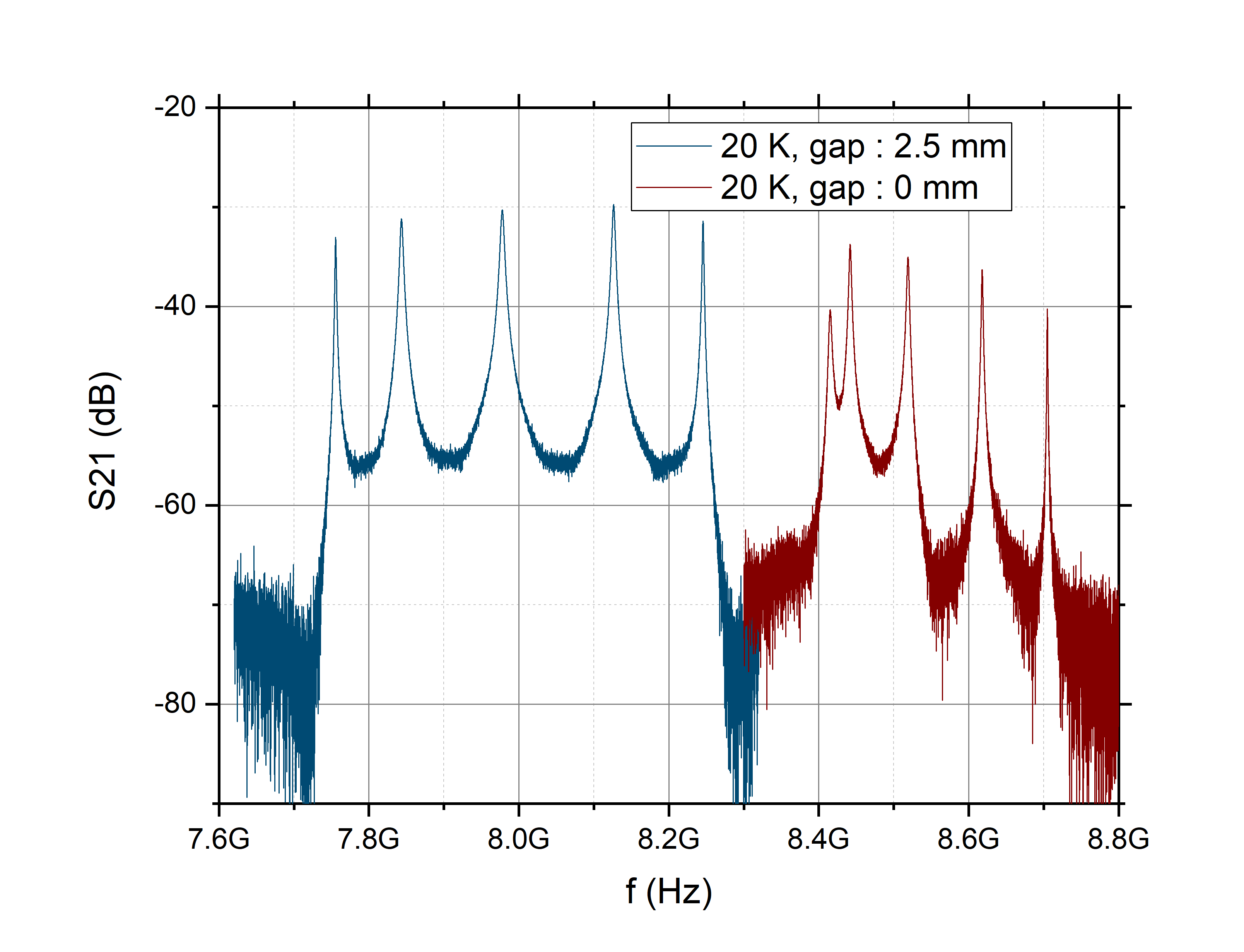}
	\end{minipage}
    \hfill
	\begin{minipage}[t]{0.49\linewidth}
		\centering 
		\includegraphics[width=01\textwidth]{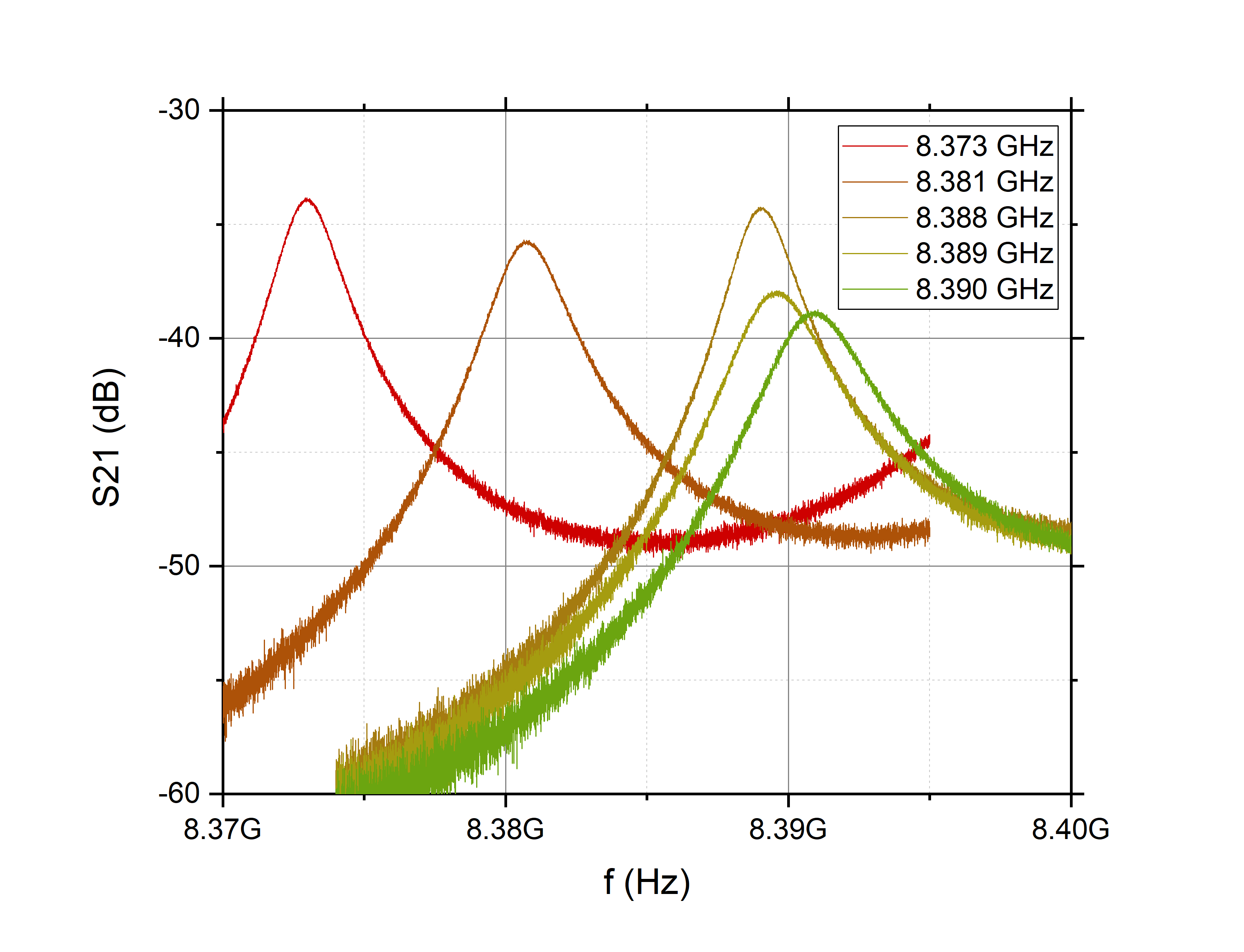}
	\end{minipage}
	\caption{Tuning range measured with the cavity embedded in the sliding structure at $20$~K with maximum gap size (blue) and without gap (red) (left). Zoom in on the first cavity peak (axion mode) for a selected set of cavity openings, demonstrating the minimum step size achieved for this measurement.}
	\label{fig:tuning20K}
\end{figure}

Fig.~\ref{fig:tuning20K} shows the tuning results at $20$~K. It was demonstrated that the cavity can be tuned in a cryogenic environment without performing a warm-up achieving a tuning range of $667$~MHz. This result confirms the tuning range predicted from the measurement in liquid nitrogen described before. Besides the tuning range, a high scanning resolution is important to investigate all possible frequencies while a haloscope is data-taking. Therefore, frequency steps in the range of the 3 dB points of the resonant frequency should be realised. The frequency step size required to achieve this depends on the cavity properties: the unloaded quality factor ($Q_0$) and the coupling of the port. Assuming critical coupling and a $Q_0$ of $27000$ (as measured for this specific cavity, see section~\ref{sec:PROOFofPri}) we aim for a step size of $0.6$~MHz ($8.4$ \ GHz$/13500$) per step.

The described set-up reached a minimum step size of $3$~MHz while the rotation was applied, turning an adjustment screw by hand; see Fig.~\ref{fig:tuning20K} (right). Using a motor turning the adjustment screw at ambient conditions will lead to higher precision. Furthermore, high-precision gears and gear racks can help to decrease the frequency step size. To maximise the detection rate during tuning, over-coupling (i.e. $\beta = 2$) instead of critical coupling is more advantageous, cf.\cite{Kim_2020}. In that case, the minimum required step size required for our cavity increases to $0.9$~MHz.

\begin{figure}[htb]
		\centering 
		\includegraphics[width=0.7\textwidth]{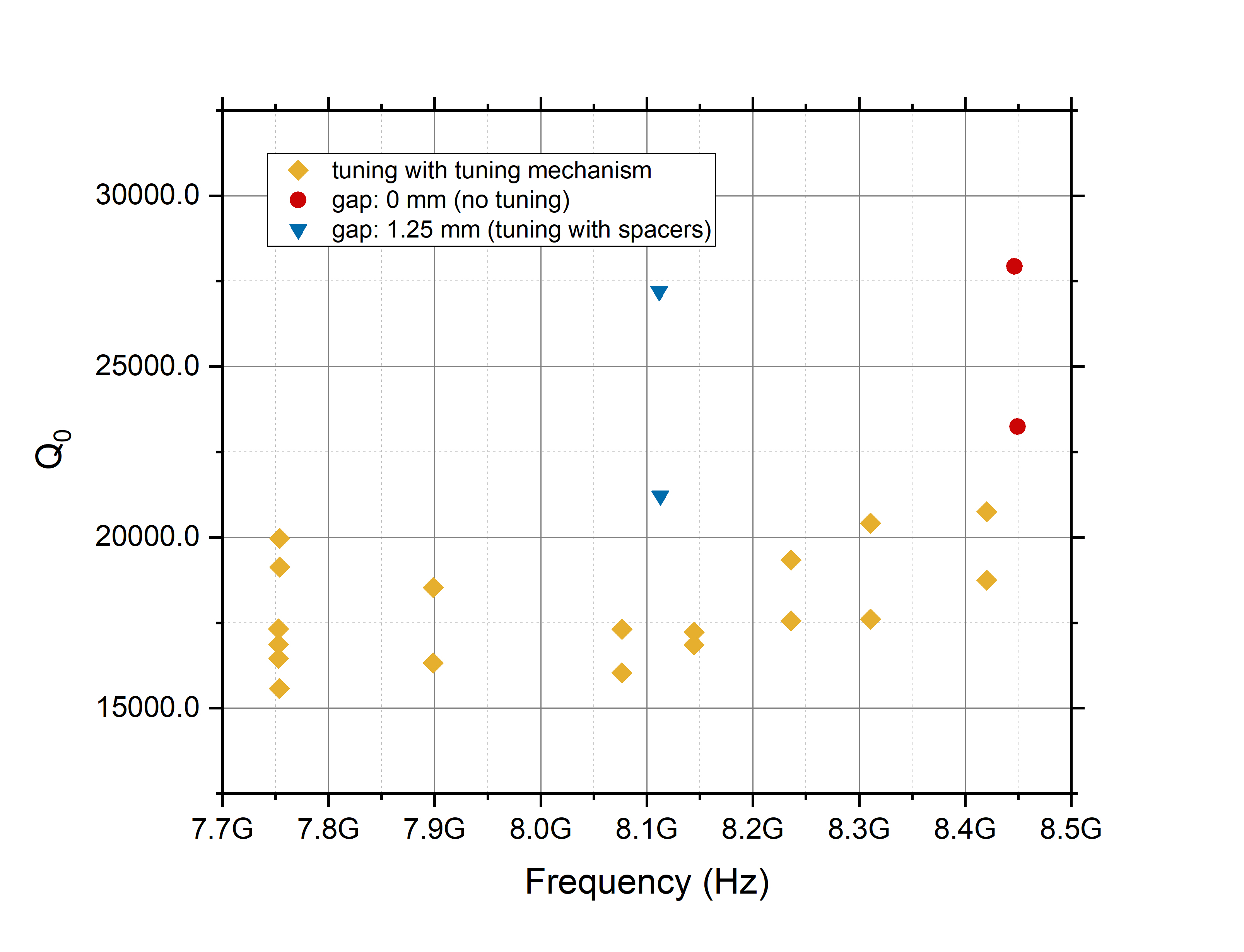}

	\caption{Unloaded quality factor ($Q_0$) of the cavity for different gap sizes generated by the tuning mechanism (yellow), by spacers (blue) and without tuning (red) measured at cryogenic temperatures below $20$~K.}
	\label{fig:tuning_QfactorCryo}
\end{figure}

The cavity must maintain a high-quality factor during tuning. Therefore, $Q_0$ of the cavity was observed while changing the resonant frequency with the tuning mechanism and compared with the results obtained with the spacers, described in section~\ref{sec:PROOFofPri}. The result of the unloaded quality factor determination for different resonant frequencies of the cavity is illustrated in Fig.~\ref{fig:tuning_QfactorCryo}.

From the simulation in section~\ref{ssec:Simulation_parallel_cavities} we expect a quality factor of about $27000$ at cryogenic temperatures which does not degrade too much for a gap of $1.25$~mm, cf. Fig.~\ref{fig:VC_verticalcuttunigresults_CST} (centre). Fig.~\ref{fig:tuning_QfactorCryo} shows that this is hard to achieve. The blue triangles mark the measurements done with a spacer (gap$ = 1.25$~mm) and the red points the cavity bolted together with screws (gap $= 0$~mm). The measurements of those configurations were already discussed in section~\ref{sec:PROOFofPri}, where we observed different quality factor values, during repeated assembly cycles. Only considering the maximum $Q_0$ measured for both spacer configurations (gap$ = 0$ and $1.25$~mm) the results are in good agreement with the simulations from section~\ref{ssec:Simulation_parallel_cavities} for parallel cavities. The lower values for $Q_0$ for the same spacer configurations are attributed to misalignments. The simulations of section~\ref{ssec:MisalingmentSim} showed how sensitive the cavity is to certain misalignment directions. The results underline the importance of a good alignment mechanism for the tuning, as a good-aligned setup is already hard to achieve with spacers.

The yellow triangles are the results obtained when changing the resonant frequency with the tuning mechanism described in section~\ref{ssec:sliding mechanism}. The quality factor is for all measurements below the values obtained with spacers and varies for the same resonant frequency. The maximum decrease in quality factor measured with the tuning mechanism compared to its optimal values for no gap ($Q_0$ about $27000$) is about $45$~$\%$. These losses are also attributed to misalignments of the two cavity halves, which leads to the conclusion that the alignment mechanism still needs to be improved and tolerances for the production of future assemblies need to be reduced.

\section{Conclusion and outlook}
\label{sec:Conclusions}

We have presented a concept to tune rectangular cavities for haloscope searches mechanically.
The concept foresees cutting cavities along a symmetry plane of the field lines. We have qualitatively validated this mechanism in practice and simulation. Assuming that the cavities are kept parallel, the simulations in section~\ref{ssec:Simulation_parallel_cavities} predict a tuning range of up to $1.1$~GHz ($12.6$~$\%$) with nearly no degradation in the figure of merit ($Q_0V^2C^2$) over the whole range.

The cavity could be tuned in a range of $667$~MHz $\sim9.5$~$\%$ at $20$~K using the sliding mechanism outlined in section~\ref{ssec:sliding mechanism}. The available bore size limited the maximum gap size ($2.5$~mm) and hence the tuning range for the presented configuration. The obtained unloaded quality factor varied between ($1.5-2.1$)~$\times10^4$ while tuning. From the simulations in section~\ref{ssec:Simulation_parallel_cavities} higher quality factors were expected. As can be observed from the simulations in section~\ref{ssec:MisalingmentSim}, the misalignment of the two cavity halves contributes
to the losses (up to $45$~$\%$) in this assembly. However, there could be other sources of losses like manufacturing tolerances or the electrical conductivity of the housing material. Previous measurements in section~\ref{sec:TUNING} showed that greater values of $Q_0$ ($2.7\times10^4$ at $7$~K, gap$ = 1.25$~mm) are possible for parallel cavity-halves (separated by spacers). This number is consistent with the parallel cavity-half simulations shown in section~\ref{ssec:Simulation_parallel_cavities}.

Moreover, misalignment can cause a $60$~$\%$ decrease in form factor C; see Fig.~\ref{fig:VC_Misalignment_CST}. This is particularly relevant for cavity designs with subcavities connected by irises, as the misalignment alters the ideal operating mode conditions. Examining the same misalignment for a long single cavity, merely a minor deterioration of the form factor ($0.66$ to $0.63$) is discernible.

The minimum step size achieved for this mechanism was $3$~MHz. To have continuous tuning steps overlapping in their $3$~dB range a step size of less than $0.9$~MHz ($3$~$\mu$m) is required. To improve the step size of the assembly, high-precision gears and a motor-driven adjustment of the tuning rod are required. 

The state of the art of various current tuning methods for different axion groups and the outcomes of this effort were displayed in Table~\ref{tab:tuningrefs} of section~\ref{sec:TUNING}. The mechanical tuning presented in this work is a competitive way to tune an axion haloscope because it can achieve the same tuning percentage as the ADMX Sidecar - Run B while also having a larger form factor and quality factor. However, it is necessary to make improvements to the current tuning design to produce a smaller step size and handle the misalignment more effectively. Further studies using an improved gear mechanism, combined with a mechanism that moves the coupler are ongoing in the group.

\section*{APPENDIX}
\subsection*{A. Misalignment due to coaxial ports}
\label{App:A_Misalignment_coaxial_ports}

In this work, an additional misalignment study has been carried out on the haloscope with vertical cut tuning. The results obtained in section~\ref{ssec:Simulation} are based on a system in which the coaxial ports are screwed to one of the two halves, causing a small misalignment or asymmetry in the total system. In this case, the behaviour of the structure has been analysed in simulation by positioning the coaxial ports centred in the middle of the opening of the structure (gap$/2$).

\begin{figure}[htb]
    \centering
    \begin{minipage}[t]{0.4\linewidth}
        \includegraphics[width=1\textwidth]{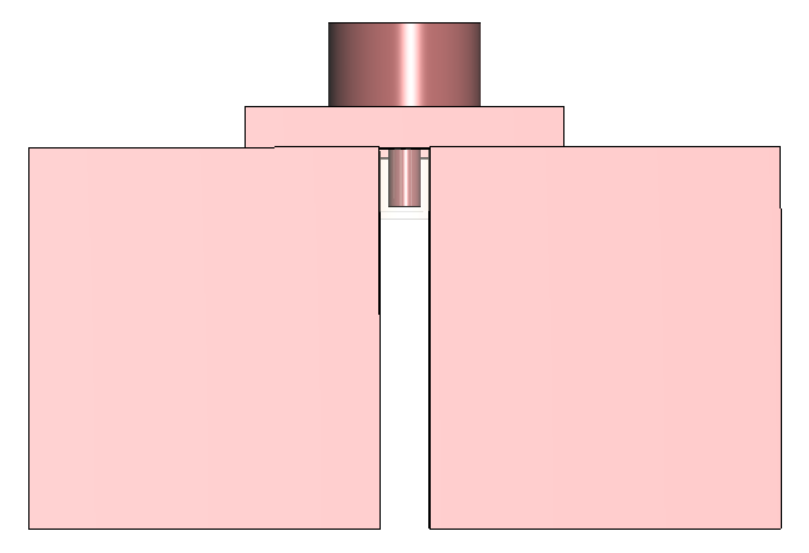}	
    \end{minipage}
    \quad \quad
    \begin{minipage}[t]{0.4\linewidth}
        \centering
        \includegraphics[width=1\textwidth]{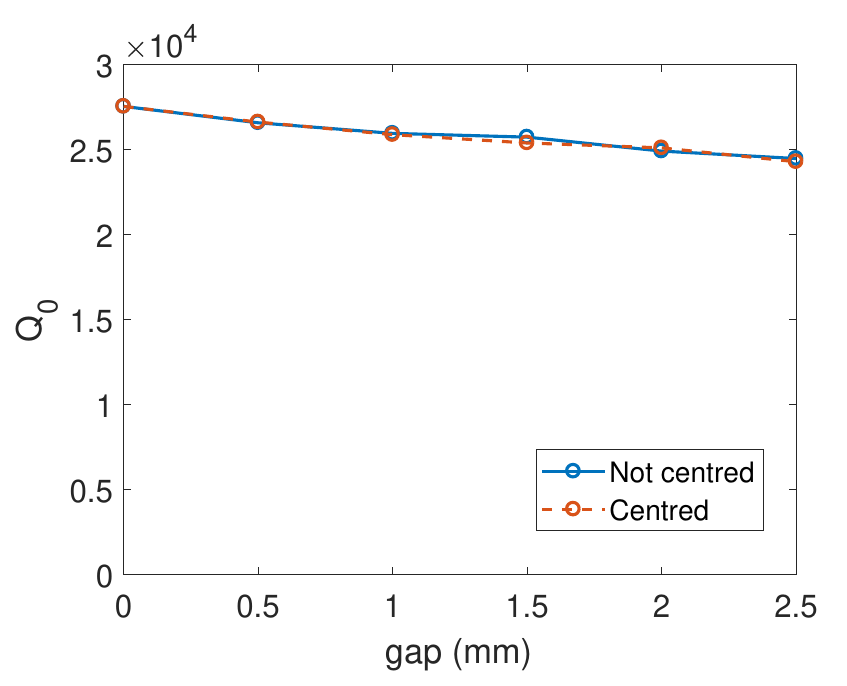}
    \end{minipage}
    \caption{(Left) 3D model of the vertical cut structure with gap$ = 2$~mm applying symmetry with the coaxial ports laying in the middle of the gap (centred at gap$/2 = 1$~mm), and (Right) quality factor results from CST simulations for this structure without (blue solid line) and with (red dashed line) centring the coaxial ports.}
    \label{fig:VC_CoaxPortsCentred}
\end{figure}

In Fig.~\ref{fig:VC_CoaxPortsCentred} the results obtained in quality factor with a range of openings from gap$ = 0$ to $2.5$~mm are depicted.
We can therefore conclude that this misalignment has practically no effect on this parameter and can be ignored in our system.

\subsection*{B. Misalignment in a long single cavity}
\label{App:B_Misalignment_long_single_cavity}

In addition, the simulation of an individual cavity with the same dimensions of width, height and total length as the vertical cut multicavity (simply eliminating the inductive irises inside) has been carried out to observe the behaviour in form factor with angular misalignment in the $y-$axis (this is $\theta_y$), which is the most relevant effect studied for this parameter (see Figure~\ref{fig:VC_Misalignment_CST} (top-left corner plot)). The width of the single cavity is $22.86$~mm, the width is $10.16$~mm, and the length is $136.36$~mm.

Applying a misalignment of $\theta_y=0^o$ to $0.6^o$, a range of detriment form factor values from $0.659$ to $0.634$ was obtained, confirming that the effect is not high for individual cavities. Thus, we can conclude that the low form factor results obtained for the vertical cut haloscope with angular $y-$axis misalignment are due to the asynchrony of the individual frequencies of the subcavities of which it is composed.

\tiny
 \keyFont{ \section{Keywords:} axion, haloscope, tuning, cryogenics, dark matter} 

\section*{Conflict of Interest Statement}

The authors declare that the research was conducted in the absence of any commercial or financial relationships that could be construed as a potential conflict of interest.

\section*{Author Contributions}

JG: Writing–review and editing, Writing–original draft, Investigation. JG-B: Writing–review and editing, Writing–original draft, Investigation. SA: Investigation, Writing–review and editing. SC: Writing–review and editing, Supervision. WW: Writing–review and editing, Supervision. BD: Writing–review and editing, Writing–original draft, Supervision, Funding acquisition.

\section*{Funding}

The authors declare that financial support was received for the research, authorship, and/or publication of this article. This work was funded by the European Research Council under grant ERC-2018-StG-802836 (AxScale project). JG is also funded by the “Physics Beyond Collider” initiative. This work has also been funded by MCIN/and by “ERDF A way of making Europe,” under grants PID2019-108122GB-C33 and PID2022-137268NB-C53. JG-B thanks the grant FPI BES-2017-079787, funded by MCIN/AEI/10.13039/501100011033 and by “ESF Investing in your future.”

\section*{Acknowledgments}
This work was performed in the RADES group and we acknowledge the contribution and invaluable help of its members. 
Additionally, we would like to thank the entire CERN cryolab team, especially Agostino Vacca for helping to modify the mechanical components of the tuning mechanism.


\section*{Data Availability Statement}
Simulation data (3D model of the vertically cut cavity in ``.stp'' format) as well as plot data for the misalignment studies are available at the \href{https://metastore.mpcdf.mpg.de/organization/rades}{MPCDF MetaStore}. Further data is available upon request.

\bibliographystyle{Frontiers-Harvard} 
\bibliography{ref_tuning.bib}


\end{document}